\documentclass[twocolumn,showpacs,amsmath,amssymb,superscriptaddress,nofootinbib,floatfix]{revtex4-1}
 
\usepackage{graphicx}
\usepackage{multirow}
\usepackage{tabularx}
\usepackage{makecell}
\usepackage{booktabs}

\usepackage{paralist}
\usepackage{subfigure}

\usepackage[colorlinks=true, linkcolor=BrickRed, citecolor=Blue, urlcolor=Blue, filecolor=Blue]{hyperref}
\usepackage[dvipsnames,table]{xcolor}

\def\nn{\nonumber}
\newcommand{\be}{\begin{equation}}
\newcommand{\ee}{\end{equation}}
\newcommand{\bea}{\begin{eqnarray}}
\newcommand{\eea}{\end{eqnarray}}
\newcommand{\beaa}{\begin{eqnarray*}}
\newcommand{\eeaa}{\end{eqnarray*}}
\newcommand{\ba}{\begin{array}}
\newcommand{\ea}{\end{array}}
\newcommand{\bi}{\begin{itemize}}
\newcommand{\ei}{\end{itemize}}
\newcommand{\ben}{\begin{enumerate}}
\newcommand{\een}{\end{enumerate}}

\newcommand{\p}{\partial}

\begin{document}

\title{Reannihilation of self-interacting dark matter}

\author{Tobias Binder}
\email{tobias.binder@theorie.physik.uni-goettingen.de}
\affiliation{Institute for Theoretical Physics, Georg-August University G\"ottingen, Friedrich-Hund-Platz 1, G\"ottingen, D-37077 Germany
}
\author{Michael Gustafsson}
\email{michael.gustafsson@theorie.physik.uni-goettingen.de}
\affiliation{Institute for Theoretical Physics, Georg-August University G\"ottingen, Friedrich-Hund-Platz 1, G\"ottingen, D-37077 Germany
}
\author{Ayuki Kamada}
\email{akamada@ibs.re.kr}
\affiliation{Center for Theoretical Physics of the Universe, Institute for Basic Science (IBS), Daejeon 34051, Korea
}
\author{Stefan Marinus Rodrigues Sandner}
\email{stefan.rodrigues@physik.uni-hamburg.de}
\affiliation{Department of Physics, University of Hamburg, Jungiusstra{\ss}e 9, D-20355  
Hamburg, Germany}
\author{Max Wiesner}
\email{mwiesner@th.physik.uni-bonn.de}
\affiliation{Bethe Center for Theoretical Physics and Physikalisches Institut der Universit\"at
Bonn, Nussallee 12, 53115 Bonn, Germany
}

\date{\today}

\begin{abstract} 
We explore the phenomenology of having a second epoch of dark matter annihilation into dark radiation long after the standard thermal freeze-out.
Such a hidden \emph{reannihilation} process could affect visible sectors only gravitationally.
As a concrete realization we consider self-interacting dark matter (SIDM) with a light force mediator coupled to dark radiation.
We demonstrate how resonantly Sommerfeld enhanced cross sections emerge to induce the reannihilation epoch.
The effect is a temporally local modification of the Hubble expansion rate and we show that the Cosmic Microwave Background (CMB) measurements --- as well as other observations --- have a high sensitivity to observe this phenomenon.
Special attention is given to the model region where late kinetic decoupling and strong self-interactions can alleviate several small-scale problems in the cold dark matter paradigm at the same time.
Interestingly, we find that reannihilation might here also simultaneously lower the tension between CMB and low-redshift astronomical observations of $H_{0}$ and $\sigma_{8}$.
Moreover, we identify reannihilation as a clear signature to discriminate between the phenomenologically otherwise almost identical vector and scalar mediator realizations of SIDM.
\end{abstract}

\maketitle

\section{Introduction}
\label{sec:intro}
The cosmological $\Lambda$CDM model has been very successful in describing the large-scale structures of the Universe. Its cold dark matter (CDM) ingredient consists of a collisionless matter component that enables to fit the observed anisotropies in the Cosmic Microwave Background (CMB)~\cite{Ade:2015xua} and to explain their evolution to form structures such as galaxies. Despite these successes, there are potential tensions between some of its predictions and observations (see, e.g., Ref.~\cite{Bullock:2017xww}). On dwarf galactic scales, there is the ``missing satellite problem'' of too few discovered satellite galaxies around the Milky Way~\cite{Moore:1999nt, Kravtsov:2009gi} and the ``core-cusp problem'' of too cored, rather than cuspy, dark matter (DM) density profiles~\cite{Moore:1999gc, deBlok:2009sp} when compared to predictions in CDM setups. The ``too-big-to-fail problem''~\cite{Vogelsberger:2012ku} tries to sharpen these arguments by pointing out that in particular the biggest satellites in simulations, which should not fail to form stars and not escape detection, have DM density profiles more concentrated than those observed~\cite{BoylanKolchin:2011de, BoylanKolchin:2011dk}.
The situation is however complicated by the fact that uncertain feedbacks from baryonic processes can be expected to play a role on sub-galactic scales~\cite{Sawala:2015cdf, Dutton:2015nvy, Wetzel:2016wro}. Nevertheless, it has been claimed in Refs.~\cite{Kamada:2016euw, Creasey:2016jaq, Robertson:2017mgj} that current state-of-the-art hydrodynamic simulations~\cite{Oman:2015xda} --- taking into account baryonic feedbacks in CDM setups --- have not been able to predict the observed diversity of rotation curves in dwarf galaxies (see, however, e.g., Ref.~\cite{2018MNRAS.473.4392S} for a possible explanation in CDM setups).

A possible way to address the missing satellite problem is to keep the DM particles in kinetic equilibrium with a relativistic species until the Universe cooled down to keV temperatures.
This would lead to dark acoustic oscillations of density fluctuations below sub-Mpc scales and consequently to the suppression of the abundance of satellite galaxies~\cite{Boehm:2000gq, Aarssen:2012fx,CyrRacine:2012fz, Kamada:2013sh, Boehm:2014vja, Schewtschenko:2014fca, Cyr-Racine:2015ihg, Vogelsberger:2015gpr, Schewtschenko:2015rno, Binder:2016pnr}.
A possibility for addressing the core-cusp problem is to have a large DM self-scattering cross section of the order of $\sigma/m_{\text{DM}} \sim 1$\,cm$^{2}$g$^{-1}$ on sub-galactic scales (with a typical rotation velocity $v_{0} \sim 30$\,km/s)~\cite{Spergel:1999mh}.
This cross section must presumably decrease with velocity to allow for a smaller impact on galaxy cluster scales (with $v_{0} \sim 10^{3}$\,km/s) where no deviations from CDM predictions are observed~\cite{Kaplinghat:2015aga}. Such strongly self-interacting dark matter (SIDM) particles would also be more sensitive to gravitational feedback from baryons, enabling them in addition to explain the diversity in rotation curves of dwarf galaxies~\cite{Kamada:2016euw, Creasey:2016jaq}.

It is interesting that all this phenomenology can arise naturally in simple three-particle models, where a light force mediator ($\phi$) induces both the desired strong DM ($\chi$) self-interaction and late DM kinetic decoupling from a thermal background particle ($l$).
The new force mediator could be either a vector or a scalar boson, both giving Yukawa potentials with proper velocity dependent $\chi$-$\chi$ scattering.
These SIDM setups can then produce the observed DM abundance via standard thermal freeze-out.

It has recently been shown that these particular types of SIDM models are strongly constrained if the mediator dominantly decays to visible standard model (SM) particles:
the vector mediator setup is in tension with indirect detection experiments and CMB observations~\cite{Bringmann:2016din, Cirelli:2016rnw, Robertson:2017hdw}; 
and most of the parameter space for scalar mediator setups are ruled out by direct detection bounds and big bang nucleosynthesis (BBN) data~\cite{Kahlhoefer:2017umn}.

A still perfectly valid setup  exists if the dark sector is essentially closed.
That is, with the mediator being a singlet under the SM gauge groups, the above constraints are clearly avoided. Meanwhile the DM self-interaction properties all remain and the thermal DM freeze-out would occur from a dark radiation background.

In this work, we investigate a novel cosmological signature of these SIDM models to discriminate between vector and scalar boson mediator setups --- in the case of a fully closed dark sector.
With a light mediator present, Sommerfeld enhanced cross sections are expected with particularly strong enhancements possible for small DM particle velocities. For $s$-wave annihilation, exclusively present in the vector mediator case, a \emph{second epoch of annihilation}~\cite{Dent:2009bv,Feng:2010zp,Zavala:2009mi,vandenAarssen:2012ag} can occur after DM kinetically decouples from its thermal background.
Here, we explore for the first time the cosmological consequences of such a reannihilation period in more detail by setting up the required coupled Boltzmann and Hubble expansion equations.
We show how this impacts the Hubble parameter significantly enough to be constrained by existing CMB data.
In the SIDM region, we interestingly find that reannihilation allows to ameliorate discrepancies in the CDM paradigm between CMB and low-redshift astronomical measurements of the Hubble rate and $\sigma_8$ \cite{Riess:2016jrr,Battye:2014qga, Salvati:2017rsn, Berezhiani:2015yta, Enqvist:2015ara, Lesgourgues:2015wza, Chudaykin:2016yfk, Poulin:2016nat}.

\medskip
The article is structured as follows.
In Section~\ref{sec:rangeneral} we review the closely degenerate phenomenology of the vector and scalar mediator SIDM setups.
In Section~\ref{sec:reannhilation}, analytic estimates and the formalism to describe reannihilation are set up.
A scan over model parameters are performed in Section~\ref{sec:analysis} and the regions where reannihilation can happen are discussed.
We investigate the impact on cosmology and constraints from CMB observation in Section~\ref{sec:CMB} for two different scenarios: reannihilation taking place before and after recombination.

\section{Light mediator setups and their phenomenology}
\label{sec:rangeneral}
In this work, we will consider two effective three-particle models, each of them having four free parameters in their Lagrangians
\bea
\mathcal{L}_{V} &\supset& g_{\chi} \bar{\chi} \gamma^{\mu} \chi \phi_{\mu} + g_{l} \bar{l} \gamma^{\mu} l \phi_{\mu} \,, \label{eq:lagrvec} \\
\mathcal{L}_{S} &\supset& g_{\chi} \bar{\chi} \chi \phi + g_{l} \bar{l}l \phi \label{eq:lagrsca} \, , 
\eea 
where $g_{\chi}$ and $g_{l}$ are the coupling constants.
The DM particle $\chi$  is a spin $1/2$ Dirac fermion with mass $m_{\chi}$ and the vector $\phi_{\mu}$ or scalar $\phi$ mediator has a mass of $m_{\phi} \ll m_{\chi}$.
The dark-radiation background particle $l$ has spin $1/2$ and is considered to be massless.

Assuming that $l$, $\phi$, and $\chi$ form a dark sector, which effectively decouples from the SM plasma, leads to an additional free parameter, namely, the temperature ratio:
\be
r \equiv \frac{T_{l}}{T_{\gamma}} \, ,
\ee
where $T_{l}$ is the dark radiation temperature and $T_{\gamma}$ is the SM photon temperature.
Fixing this temperature ratio at a particular time, e.g.,  at the temperature $T_{\gamma}^{\text{BBN}} = 1$\,MeV, its further temperature dependence is given from entropy conservation as
\be
r(T_{\gamma}) = r_{\text{BBN}} \left( \frac{g_{s}(T_{\gamma})}{g_{s}(T_{\gamma}^{\text{BBN}})} \right)^{1/3} \, , \label{eq:revolution}
\ee
where $g_{s}$ is the SM entropy degrees of freedom and we assume that entropy production in the dark sector can be ignored after DM chemically decouples.
In this work we choose, unless quoted differently, $r_{\text{BBN}} = 0.5$, which turns into $r \simeq 0.35$ after electron-positron decoupling and is compatible with current BBN constraints~\cite{Nollett:2014lwa, Hufnagel:2017dgo}.
Such ratios are achieved if the dark sector kinetically decouples from the SM plasma above a temperature of $T_{\gamma} \simeq 40$\,GeV.
For this work it is however not required to specify the coupling to the SM leading to kinetic equilibration between the two sectors.
Temperature ratios of this order could also be achieved by some inflationary models.

In the rest of this section we will highlight the similarities and differences between the two models and present the phenomenological results to be used in subsequent sections.

\subsection{Velocity dependent self-interactions}
DM self-interactions lead to an iso-thermal DM velocity distribution in the inner region of halos.
If the self-scattering cross section is of the order of $\sigma / m_{\text{DM}} \sim 1$\,cm$^{2}$g$^{-1}$ the DM density distribution in dwarf galaxies is characterized by a kpc-size core.
This mechanism enables to mitigate the core-cusp and the too-big-to-fail problems~\cite{Elbert:2014bma}.
SIDM alone, however, cannot explain the observed diversity of dwarf galaxy rotation curves since it changes the density profile universally among similar-size halos.
Rotation velocities in the inner region are observed to differ by up to a factor of $\simeq 4$ among halos with a similar rotation velocity in the outer region. This diversity was not predicted by state-of-the-art hydrodynamic simulations taking into account baryonic feedbacks like galaxy formation and supernova explosions~\cite{Oman:2015xda}.
A key observation is however that the SIDM profile is quite sensitive to the presence of the baryonic bulge and disk in the inner part of a galaxy.
An iso-thermal DM velocity distribution is determined by the total gravitational potential, which in the inner region can be dominated by the galaxy's baryonic content.
Together with the measured baryon distribution, SIDM is able to address the observed diversity in dwarf galaxy rotation curves~\cite{Kamada:2016euw, Creasey:2016jaq}.

A thermalized DM halo, on the other hand, may be incompatible with observations of galaxy clusters.
Its distribution is virtually spherical, but a strong lens system prefers a sizable ellipticity of the lens galaxy cluster~\cite{MiraldaEscude:2000qt}.
While the projection effect in the lens analysis is subject to caveat, the constraint would be as severe as $\sigma / m_{\text{DM}} \lesssim 0.1$\,cm$^{2}$g$^{-1}$~\cite{Peter:2012jh}.
A merging cluster system like a bullet cluster also provides a good test for SIDM.
The reported tight constraint is $\sigma / m_{\text{DM}} \lesssim 0.7$\,cm$^{2}$g$^{-1}$~\cite{Randall:2007ph}; otherwise a sizable amount of DM mass evaporates from the subclusters during the collision and the resultant system is incompatible with the observed mass-to-light ratios.
One may have to be careful about the uncertainty in the unconstrained initial condition of the system.
Although it is too early to conclude (see, e.g., Ref.~\cite{Tulin:2017ara} for a comprehensive summary), the velocity dependence may have to be introduced into the self-scattering cross section to reduce the effects of SIDM in galaxy clusters while keeping a sizable cross section in dwarf galaxies.
The desired velocity dependence can naturally be realized in both the light vector and and scalar mediator setups of Eqs.~\eqref{eq:lagrvec} and \eqref{eq:lagrsca} \cite{Tulin:2013teo, Tulin:2012wi}.

The averaged self-scattering cross section in a thermal DM halo with a characteristic velocity $v_{0}$ can be computed from
\be
(\sigma_{T})_{v_{0}} = \frac{4 \pi}{(\sqrt{2 \pi} v_{0})^{3}} \int \sigma_{T} \, e^{- v_{\text{rel}}^{2} / (2 v_{0}^{2})} \, v_{\text{rel}}^{2} \, \text{d} v_{\text{rel}} \, ,
\ee
where $\sigma_{T}$ is the transfer cross section:
\be
\sigma_{T} \equiv \int \text{d} \Omega \, (1-\cos \theta) \frac{\text{d} \sigma_{\chi \chi\rightarrow \chi \chi}}{\text{d} \Omega} \, .\label{eq:transfercrx}
\ee
For the Yukawa-potential scattering, induced by our light mediators, we will use the ETHOS fitting functions for $\sigma_{T}$ in the classical regime ($m_{\chi} v_{\text{rel}} \gtrsim m_{\phi}$) 
--- as they are provided in Eqs.~(45) and (46) of Ref.~\cite{Cyr-Racine:2015ihg} and originally proposed in Ref.~\cite{Tulin:2013teo}.
In the parameter region where $s$-wave scattering is dominant (quantum-resonant regime), we will use the analytic expression provided in Eq.~(A5) of Ref.~\cite{Tulin:2013teo}.
We assume DM to be symmetric and average over particle and antiparticle scattering contributions as suggested in Ref.~\cite{Cyr-Racine:2015ihg}.
In Section~\ref{sec:analysis} we use these expressions to find the model parameter region where ($ \sigma_{T})_{30 \, \text{km/s}}/m_{\chi} \in [0.1,10] \, \text{cm}^{2} \text{g}^{-1}$ (sizable self-scattering cross section on dwarf galactic scales).

We remark that neither the ETHOS fitting functions nor the numerical solution of the scattering amplitude in Ref.~\cite{Tulin:2013teo} account for the correct quantum statistics in their computation of $\sigma_{T}$.
They rely on classical assumptions like the distinguishability of DM particles.
It is hard to realize proper quantum corrections in SIDM $N$-Body simulations, but when adopting a classical treatment it is at least important to keep track on the expected theoretical uncertainties.
In Appendix~\ref{app:ssxs} we compare the commonly adopted classical approximation of $\sigma_{T}$ to the proper quantum mechanical treatment~\cite{Kahlhoefer:2017umn} for all the scattering possibilities of our mediator setups. We find that there are up to factor two corrections on dwarf galactic scales between these two approaches, however, they have no relevant impact on our results.

\subsection{Dark acoustic oscillations}
The coupling between non-relativistic DM particles and radiation leads to competition between gravity and radiative pressure.
The pressure effect is strong during kinetic equilibrium between DM and relativistic particles, leading to \emph{dark acoustic oscillations} of the DM density perturbations inside the causal horizon.
Therefore, matter density fluctuations can only grow significantly after DM kinetically decouples.
This leads to the fact that the resultant matter power spectrum is suppressed on length scales shorter than the Hubble horizon distance at kinetic decoupling.
The minimal protohalo mass (or cutoff mass) below which the halo mass function is suppressed can be estimated by the mass inside a sphere with the radius of the Hubble horizon at the time of DM kinetic decoupling:
\bea
M^{\text{cut}} &\equiv& \rho_{m} \frac{4 \pi}{3}\left( \frac{1}{H} \right)^{3} \Big|_{\text{{DM \bf{k}}inetic {\bf{d}}ecoupling}} \nonumber \\ 
&=& 2.2 \times 10^{8} r^{3}_{\text{kd}} \left( \frac{1 \,\text{keV}}{T_{l}^{\text{kd}}} \right)^{3} \, M_{\odot} \,.
\label{eq:mcutgleichung}
\eea
Here, $\rho_{m}$ is the total matter density and $H$ is the Hubble expansion rate.
We see that a kinetic decoupling temperature $T_{l}^{\text{kd}}$ of the order of $1$ keV leads to the suppression of the halo mass function on dwarf-galaxy masses and hence addresses the missing satellite problem.
Especially in our case of late kinetic decoupling and non-relativistic DM, this damping dominates over the free-streaming effect. The damping mechanism of dark acoustic oscillations has been extensively investigated by many authors~\cite{Boehm:2000gq, Boehm:2001hm, Chen:2002yh, Sigurdson:2004zp, Boehm:2004th, Mangano:2006mp, Serra:2009uu, Aarssen:2012fx,CyrRacine:2012fz, Kamada:2013sh, Wilkinson:2013kia, Cyr-Racine:2013fsa, Dvorkin:2013cea, Wilkinson:2014ksa, Boehm:2014vja, Schewtschenko:2014fca, Cyr-Racine:2015ihg, Vogelsberger:2015gpr, Schewtschenko:2015rno, Escudero:2015yka, Ali-Haimoud:2015pwa, Binder:2016pnr, Kamada:2017oxi} --- in part also in the context of SIDM. For a classification of DM models leading to late kinetic decoupling we refer readers to Ref.~\cite{Bringmann:2016ilk}.

In our setups both the mediator $\phi$ and the fermionic particle $l$ can act as pressure sources leading to dark acoustic oscillations. In the parameter space we will consider, the scattering between DM $\chi$ and $l$ dominates over that between $\chi$ and $\phi$.
The kinetic decoupling temperature $T_{l}^{\text{{kd}}}$ can be defined as the temperature when the Hubble expansion rate $H$ equals the momentum transfer rate $\gamma$.%
\footnote{For a more precise definition of a kinetic decoupling temperature and its matching to the non-linear cutoff in the matter-power spectrum, see Refs.~\cite{Cyr-Racine:2015ihg, Vogelsberger:2015gpr, Bringmann:2006mu, Bringmann:2009vf}.} 
In Refs.~\cite{Binder:2016pnr, Cyr-Racine:2015ihg} the momentum transfer rate is derived to be
\begin{eqnarray}
\! \! \! \gamma
\! =
\! \frac{1}{6 m_{\chi} T_{l}}\! \sum_{s_{l}} \!\int\! \frac{d^{3} \mathbf{p}_{l}}{(2\pi)^{3}}
f^{\text{eq}}_{l} 
(1 - f^{\text{eq}}_{l})
\! \! \! \int\displaylimits^{0}_{-4\mathbf{p}_{l}^{2}} \! \! \! dt (-t)
\frac{d\sigma}{dt} v_{\text{rel}} \,,
\label{eq:gammanonrel}
\end{eqnarray}
and explicit expressions of the elastic $l$-$\chi$ scattering cross section $d\sigma / dt \, v_{\text{rel}}$ and the kinetic decoupling temperature can be found in Refs.~\cite{Binder:2016pnr, Bringmann:2016ilk}.
This momentum transfer rate of $l$-$\chi$ scattering scales as $\gamma \propto T_{l}^6$ for both scalar and vector mediators.
Furthermore, both scenarios acquire the same minimal halo mass $M^{\text{cut}}$ for similar coupling constants and particle masses~\cite{Binder:2016pnr}.
The suppression of the halo mass function mainly depends on the mediator mass $m_{\phi}$ and for both models a cutoff mass around the dwarf galactic scale can be achieved for $m_{\phi} $ of the order of $1$\,MeV.

The predicted matter power spectra for scalar and vector mediators differ slightly in shape due to differences in the angular dependence of their $\chi$-$l$ scattering amplitudes~\cite{Cyr-Racine:2015ihg, Vogelsberger:2015gpr}.
As a consequence, the two models are in principle distinguishable, but in a recent study~\cite{Archidiacono:2017slj} it was shown that the differences are too small to be seen in current CMB observations.
Future observations of CMB spectral distortions, however, might be sensitive enough to discriminate models where DM is kept in kinetic equilibrium via SM photon scattering and where the DM kinetic equilibrium is kept via SM neutrino scattering~\cite{Diacoumis:2017hff}.

Recently, a combined Ly-$\alpha$ forest data analysis~\cite{Irsic:2017ixq} constrained the damping of the matter power spectrum due to the free-streaming effect of warm dark matter (WDM).
The lower limit on the WDM mass can be approximately translated into a lower limit of the kinetic decoupling temperature by equating the suppressed matter power spectra in a certain range of wavelengths.
The authors of Ref.~\cite{Huo:2017vef} found with this estimate a lower limit of $T_{l}^{\text{kd}}/r \gtrsim 1 \, \text{keV}~(0.6 \, \text{keV})$,%
\footnote{The strong (respective the weak) limit is derived from the Lyman-$\alpha$ measurements in Ref.~\cite{Irsic:2017ixq}, where a power-law assumption (respective a free floating value) is used to describe the redshift evolution of the intergalactic medium temperature.}
which results in according to Eq.~\eqref{eq:mcutgleichung} an upper limit on the cutoff mass of 
\be
M^{\text{cut}} \lesssim 2 \times 10^{8} \, M_{\odot} ~ (10^{9} \, M_{\odot}) \, .
\ee
Cutoff masses in the range $10^{7}$ to $10^{9} \, M_{\odot}$ are indicated in Fig.~\ref{fig:deg} of  Section~\ref{sec:analysis},
where $M^{\text{cut}}$ is a function of our model parameters as determined from Eqs.~(3.14) in Ref.~\cite{Binder:2016pnr}.

\subsection{Sommerfeld enhanced annihilation}
The two models differ in the leading-order cross section results of their DM annihilation channels: $\chi \bar{\chi} \rightarrow \phi \phi$ and $\chi \bar{\chi} \rightarrow l \bar{l}$.
In the vector mediator case, both processes are $s$-wave dominated and in the scalar mediator case they are $p$-wave dominated.
Generically for both models, the annihilation rate is Sommerfeld enhanced in the DM non-relativistic regime and for each particular annihilation channel the cross section factorizes into a short and a long-range contribution:
\bea
(\sigma v_{\text{rel}})_{V} &\simeq& S_{s} (v_{\text{rel}}) \sum\limits_{i} (\sigma v_{\text{rel}})^{s}_{0, i}  \, ,\\
(\sigma v_{\text{rel}})_{S} &\simeq& S_{p} (v_{\text{rel}}) \sum\limits_{i} (\sigma v_{\text{rel}})^{p}_{0, i}  \, .
\eea
The long-range force corrections are encoded in the velocity dependent Sommerfeld factor $S(v_{\text{rel}})$ multiplying \emph{universally} the tree-level cross section $(\sigma v_{\text{rel}})_{0, i}$ for each particular annihilation channel $i$.
In particular, the vector mediator model has
\bea
(\sigma v_{\text{rel}})^{s}_{0, \phi \phi} &\simeq& \frac{\pi \alpha_{\chi}^{2}}{m_{\chi}^{2}},\label{eq:sv0vecPhi} \\
(\sigma v_{\text{rel}})^{s}_{0, l \bar{l}} &\simeq&  \frac{\pi \alpha_{\chi} \alpha_{l}}{m_{\chi}^{2}}, \label{eq:sv0vecl}
\eea
where $\alpha_{\chi (l)} \equiv g^{2}_{\chi (l)} / 4 \pi$ and in the scalar mediator case the corresponding tree-level cross sections are instead $(\sigma v_{\text{rel}})^{p}_{0, i} \propto v^{2}_{\text{rel}}$.

$S(v_{\text{rel}})$ can be obtained from the two DM particles' wave-function value at the interaction point by numerically solving their Schr\"odinger equation with the potential resulting from $t$-channel exchanges of the light mediator~\cite{Hisano:2003ec, Hisano:2004ds, Hisano:2006nn, Iengo:2009xf}.
In the static limit, both mediator types induce a Yukawa potential.
It was shown in Ref.~\cite{Cassel:2009wt} that the Sommerfeld factor resulting from the Hulth\'en potential describes very accurately the numerical solution of the Schr\"odinger equation with a Yukawa potential.
The advantage of the Hulth\'en potential is that analytic solutions of $S$ for $s$- and $p$-waves are known.
The expression of $S$ for $s$-wave annihilation is given by (see, e.g., Ref.~\cite{Cassel:2009wt}) 
\be
S_{s} (v_{\text{rel}})= \frac{\pi}{\epsilon_{v}} \frac{\sinh\left(\frac{ 12 \epsilon_{v} }{ \epsilon_{\phi} \pi} \right)}{\cosh\left(\frac{ 12 \epsilon_{v} }{ \epsilon_{\phi} \pi} \right)-\cos\left[ 2\pi \sqrt{\frac{6}{ \epsilon_{\phi} \pi^{2}} - \left(\frac{6 \epsilon_{v} }{\epsilon_{\phi} \pi^{2} }\right)^{2}}\right]} \,, \label{eq:sofv}
\ee
where the two dimensionless parameters are defined as
\bea
\epsilon_{v} &\equiv& \frac{v_{\text{rel}}}{2 \alpha_{\chi}} \, ,\\
\epsilon_{\phi} &\equiv& \frac{m_{\phi}}{ \alpha_{\chi} m_{\chi}} \, .
\eea
The Sommerfeld factor in Eq.~\eqref{eq:sofv} is resonantly enhanced if the parametric condition,
\be 
\epsilon_{\phi} = \frac{6}{n^{2} \pi^{2}}  \quad  \text{with} ~ n \in \mathbb{Z}^+  \, , \label{eq:resonancecondition}
\ee
is fulfilled. The position of the $n^{\text{th}}$ ``Sommerfeld resonance'' is the same for $s$- and $p$-wave annihilation (except for $n=1$ where the resonance is absent in $S_p$).
We will refer to this equation as the ``parametric resonance condition''. 
For this work, the most important difference is that only in the case of $s$-wave annihilation, the total cross section scales as 
\be 
(\sigma v_{\text{rel}})_{V} \propto  v_{\text{rel}}^{-2} \quad \text{for}  ~ v_{\text{rel}} \lesssim  m_{\phi} / m_{\chi} \, , \label{eq:srelvec}
\ee
when $\epsilon_{\phi}$ is close to a resonance condition.
For $p$-waves, the cross section is constant in this regime and never scales stronger than $v^{-1}$.
The $v^{-2}$ feature of $(\sigma v_{\text{rel}})_{V}$ is thus only available in the vector mediator model.

The implications of the strongly velocity dependent enhancement in Eq.~\eqref{eq:srelvec} are the main part of this work.
As we will see, it can lead to a reannihilation period where the comoving DM number density significantly decreases a second time.
What is important to note is that the analytic formula of the Sommerfeld factor as given in Eq.~\eqref{eq:sofv} can violate the $s$-wave unitarity bound for low velocities when the parametric resonance condition is exactly (or almost) fulfilled.
In the numerical analyses in the subsequent sections we will therefore always use the improved analytic solution provided in Ref.~\cite{Blum:2016nrz}, accounting for a physical behavior on top of a resonance and correcting the approximate expression in Eq.~\eqref{eq:sofv} for extremely low relative velocities.
In Appendix~\ref{app:selfconsistentS}, we provide the details of this improved analytic formula and discuss the important role of saturation of $(\sigma v_{\text{rel}})_{V}$ below the unitarity limit.

\section{An Epoch of Reannihilation}
\label{sec:reannhilation}
In the previous section we put emphasis on the very similar phenomenology of the two light mediator models in Eqs.~\eqref{eq:lagrvec} and \eqref{eq:lagrsca}: they are practically identical  candidates for alleviating multiple small-scale structure formation issues in a comparable model parameter space.
In the following, we point out that even in the case of not including any couplings at all to SM particles and therefore ``hiding'' the dark sector, the \emph{impact on cosmology at late times can be significantly different}.
More precisely, we show that only in the vector mediator case a strong Sommerfeld enhancement, such as in Eq.~\eqref{eq:srelvec}, can lead to a second period of annihilation.

\begin{figure}
\centering
\includegraphics[width=0.95\columnwidth]{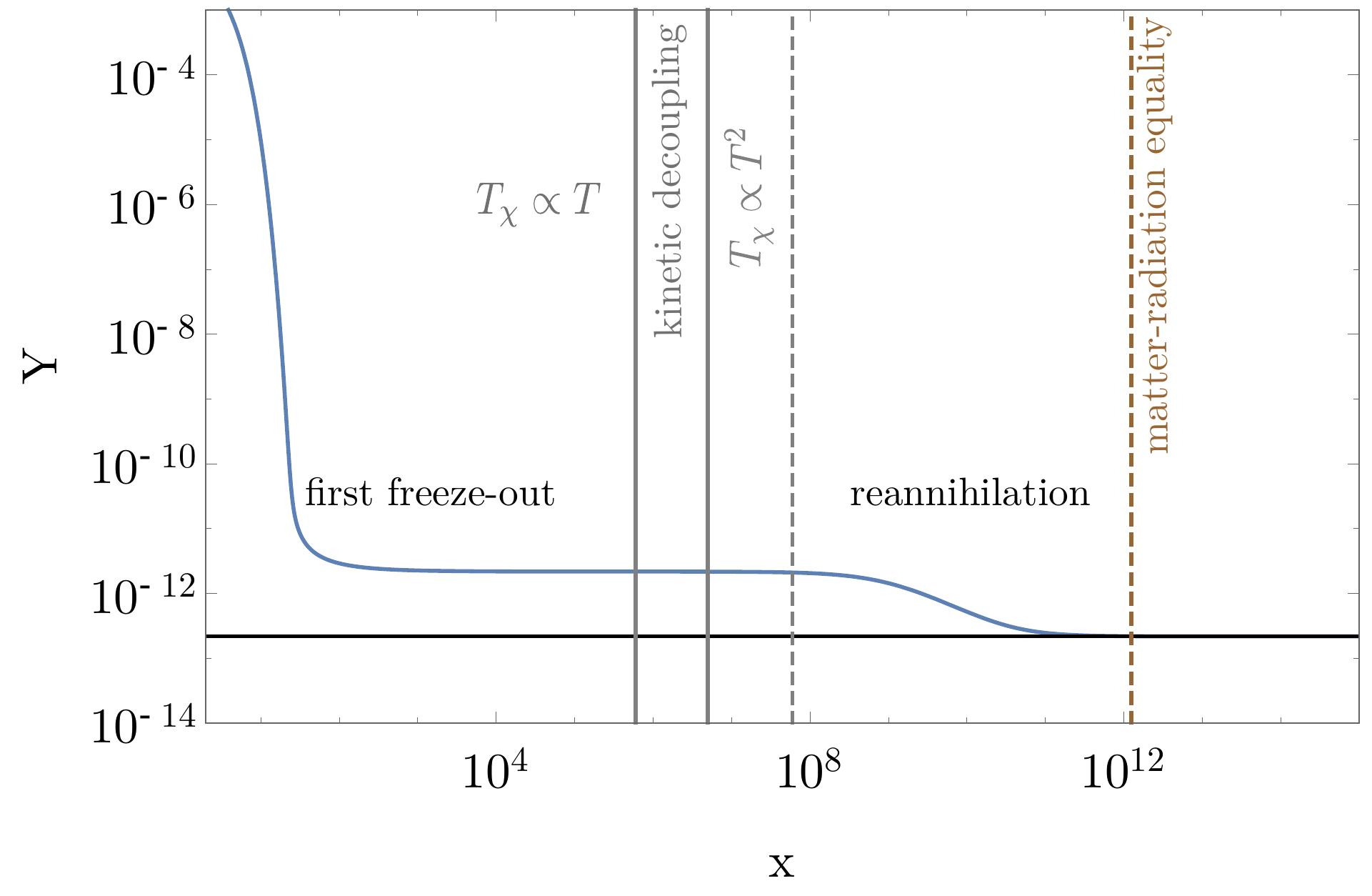}
\caption{Reannihilation process shown as a function of $x\equiv m_{\chi} /T_{\gamma}$, changing the DM co-moving number density $Y\equiv n_{\chi}/s $ by one order of magnitude.
The final abundance coincides with the correct value (black horizontal line).
Here, we have chosen the parameters 
$m_{\chi} = 1 \, \text{TeV}$, $\alpha_{\chi} = 0.007$, $m_{\phi} \simeq 1 \, \text{GeV}$ and the resonance number $n \simeq 2$ (where $m_{\phi}$ is tuned to get the correct relic density).
Between the gray solid lines kinetic decoupling happens and the scaling of the DM temperature changes from $T_{\chi} \propto T$ to $T_{\chi} \propto T^{2}$.
The dashed gray line indicates the start of reannihilation, where velocity-dependent annihilation lead to deviation from the $T_{\chi} \propto T^{2}$ scaling.
}
    \label{fig:freezeoutfull}
\end{figure}
In Fig.~\ref{fig:freezeoutfull} an example of  a reannihilation epoch is shown. After kinetic decoupling the DM abundance decreases by one order of magnitude before the time of matter-radiation equality.
The final $\chi$ relic abundance coincides with the observed CDM value, {$\Omega_{c} h^{2} = 0.1197$ (central value of ``Planck 2015 (TT+lowP)" analysis~\cite{Ade:2015xua}).
In Section~\ref{sec:estimates}, we provide analytic estimates and an intuitive understanding of when and in which region of the parameter space of the vector mediator model reannihilation can happen.

The reannihilation process necessarily starts after kinetic decoupling, as in the example of Fig.~\ref{fig:freezeoutfull}.
During this process, the evolution of the DM temperature $T_{\chi}$ does not follow the typical $T_{\chi} \propto T_{l}^{2}$ scaling for kinetically decoupled non-relativistic particles, since the Sommerfeld enhancement leads to a strongly velocity dependent annihilation cross section.
The DM particles preferably annihilate at low momenta, which leads to an increase of the DM temperature and in turn influences the annihilation rate.
It requires to go beyond the standard way of describing the DM number density evolution~\cite{Gondolo:1990dk} to cover kinetic decoupling and the impact of DM velocity dependent annihilation on the DM temperature.
In Section~\ref{sec:nummeth} we adopt the method developed in Ref.~\cite{vandenAarssen:2012ag} (and first estimated in Ref.~\cite{Feng:2010zp}) of how to deal with the number density computation in such a case correctly.
We further extend the set of equations by including the impact of the injected dark radiation on the expansion rate.
Moreover, we provide a reliability check of the method proposed in Ref.~\cite{vandenAarssen:2012ag} by solving the Boltzmann equation at phase-space density level.

\subsection{Estimates}
\label{sec:estimates}
To analytically quantify if and when DM reannihilation happens, we study the ratio between the annihilation and expansion rates:
\be
\Gamma \equiv \frac{\langle \sigma v_{\text{rel}}\rangle_{x'} \, Y }{H / s} \, , \label{eq:gammadef}
\ee
where the dimensionless form of the DM number density $n_{\chi}$ is defined as
\be
Y\equiv \frac{n_{\chi}}{s} \, . \label{eq:defnumberdensity}
\ee
Here, the SM entropy $s = g_{s} (2\pi^{2} / 45)  T_{\gamma}^{3}$ and the Hubble expansion rate,  $H \propto \sqrt{g_{\rm eff}}\,T_\gamma^2$ 
during radiation domination and $\propto T_\gamma^{3/2}$ during matter domination, are both dynamical functions of
\be 
x\equiv \frac{m_{\chi}}{T_{\gamma}} \, . 
\ee
We follow the evolution of $\Gamma$ after the first freeze-out, so that we can assume $Y$ to be constant until the start of reannihilation.
The thermally averaged cross section is a function of the DM temperature $T_{\chi}$ and can be written in the non-relativistic limit as
\be
\langle \sigma v_{\text{rel}} \rangle_{x'} = \frac{(x^{\prime})^{3/2}}{2 \sqrt{\pi}} \int_{0}^{\infty} (\sigma v_{\text{rel}}) \, e^{-x^{\prime} v_{\text{rel}}^{2} / 4} \, v_{\text{rel}}^{2} \, \text{d} v_{\text{rel}} \, , \label{eq:sigmavaveraged}
\ee
where it is a function of
\be
x^\prime \equiv \frac{m_{\chi}}{T_{\chi}} \, . 
\ee
We note that for a cross section of the form of $(\sigma v_{\text{rel}}) \propto v_{\text{rel}}^{-n}$, where we consider here only $n=0,1,2$, the thermally averaged cross section can be computed analytically and scales as $\langle \sigma v_{\text{rel}} \rangle_{x'} \propto {x'}^{-n/2}$.

To now estimate the scaling of $\Gamma$ as a function of $x$, we approximate the kinetic decoupling as an instantaneous process such that we can write
\bea
x^{\prime} = 
\begin{cases}
 \frac{m_{\chi}}{T_{l}} = \frac{x}{r} & \text{before kinetic decoupling,} \\  
 \frac{m_{\chi} T_{l}^{\text{kd}}}{T_{l}^{2}} = \frac{x^{2}}{r^{2} x^{\text{kd}}_{l}}  & \text{after kinetic decoupling.} 
\end{cases} 
\label{eq:xpTOx}
\eea
Here, $x^{\text{kd}}_{l} \equiv m_{\chi}/T_{l}^{\text{kd}}$ and the dynamical temperature ratio $r$ is defined in Eq.~\eqref{eq:revolution}.
The exact evolution of $x'$, beyond the approximation of instantaneous kinetic decoupling, is a part of the next Section~\ref{sec:nummeth}.

We provide in Table~\ref{tab:tab1} the scaling of $\Gamma(x)$ for different types of velocity dependent cross sections in the instantaneous kinetic decoupling approximation.
\begin{table*}
\begin{center}
{\renewcommand{\arraystretch}{1.2}
    \begin{tabularx}{0.7\textwidth}	{*{4}{>{\centering\arraybackslash}X|} >{\centering\arraybackslash}X}    
    			& 	\multicolumn{4}{ c }						{$( \sqrt{g_{\text{eff}}}/g_{s}) \times \Gamma \propto$} 												\\  \cline{2-5}
				\multicolumn{1}{ c | } {\multirow{2} {*}  		{\hspace{3mm} $(\sigma v_{\text{rel}}) \propto$ \hspace{3mm}} }  
    			& 	\multicolumn{2}{ c| } 	{radiation-dominated epoch}  			& \multicolumn{2}{ c }{matter-dominated epoch}  						\\  \cline{2-5}
    			& 	\centering {\bf before} kinetic decoupling  &  \centering {\bf after} kinetic decoupling 		& \centering {\bf before} kinetic decoupling  	& \hspace{3mm}	 {\bf after} kinetic decoupling 	    	\\ 
			\Xhline{4\arrayrulewidth}
    constant & $ x^{-1}$ 				& $ x^{-1}$ 					& $ x^{-3/2}$				& $ x^{-3/2}$						\\\hline
    $1 / v_{\text{rel}}$ 		& $r^{-1/2}$ $x^{-1/2}$ 		& $r^{-1}$ 					& $r^{-1/2}  x^{-1}$			& $ r^{-1}  x^{-1/2}$ 					\\ \hline
    $1 / v_{\text{rel}}^{2}$  	& $r^{-1}$ 				& $ r^{-2}  \boldsymbol x$ 			& $r^{-1} x^{-1/2}$ 			& $ r^{-2}  \boldsymbol x^{1/2} $    
    \end{tabularx}
}
\end{center}
\caption{Evolution of $\Gamma$ at different cosmological epochs and for different DM annihilation cross sections $(\sigma v_{\text{rel}})$.
Only after kinetic decoupling and with a cross section scaling as $1/v^{2}$, the ratio of annihilation over expansion rate, $\Gamma$, can grow to enter a period of reannihilation.
Sommerfeld enhanced $s$-wave annihilation features such a $1/v^{2}$ scaling.
$r$ is the ratio of dark radiation to photon temperature, as defined in Eq.~\eqref{eq:revolution}.}
\label{tab:tab1}
\end{table*}
Let us discuss some of its entries in the temporal order of the example scenario shown in Fig.~\ref{fig:freezeoutfull}.
After chemical decoupling, where $\Gamma$ drops below 1, $\Gamma$ scales as  $x^{-1}$ until the Sommerfeld factor (or the total $s$-wave annihilation cross section) starts to dominantly scale as $S \propto 1/v_{\text{rel}}$.
From this point to kinetic decoupling $\Gamma$ further decreases in the phase of the $S \propto 1/v_{\text{rel}}$ scaling, followed by a period where $\Gamma$ stays constant.
After kinetic decoupling, when $T_{\chi}$ starts to drop quickly, the $S \propto 1/v_{\text{rel}}^{2}$ scaling dominates and leads to an increase of $\Gamma$ as is highlighted by boldface $\boldsymbol x$ in the table.
When $\Gamma$ starts to approach 1 again, the DM abundance significantly decreases a second time as seen in Fig.~\ref{fig:freezeoutfull}.
The reannihilation process stops when the Sommerfeld enhancement is saturated, finally leading to $(\sigma v_{\text{rel}}) \propto \text{constant}$ and to a fast decrease of $\Gamma$ --- as can be read off from Table~\ref{tab:tab1}.
The saturation velocity depends on how parameters combine to the resonance condition, given in Eq.~\eqref{eq:resonancecondition}. For Fig.~\ref{fig:freezeoutfull} we have chosen a point slightly next to a resonance, such that the saturation effect gives the correct value of the relic density. If exactly on top of the resonance, the reannihilation process would have been longer, further reducing the abundance of DM. In general, the evolution pattern of $\Gamma$ can vary depending on the model parameters.

\medskip
An important quantity, used in the following sections, is the redshift $z_{\text{rea}}$ defined by when the DM co-moving number density changes first by $1 \, \%$ due to reannihilation.
$z_{\text{rea}}$ can be determined from the value of $\Gamma$ and a detailed derivation is provided in Appendix~\ref{app:reaestimates}. For our vector mediator model, we find that the onset of the reannihilation process is roughly given by
\be
z_{\text{rea}} \simeq 100 \times \left( \frac{r_{\text{kd}}}{0.36} \right)^{-5} \left( \frac{\alpha_{\chi}}{0.02} \right) \left( \frac{m_{\chi}}{ \text{TeV}} \right)^{-3/2} \left( \frac{m_{\phi}}{1.2 \, \text{MeV}} \right)^{4} \,, \label{eq:zrea}
\ee
where we have assumed $\alpha_{\chi} = \alpha_{l}$ and that $r$ does not change after kinetic decoupling.
From this equation it can be recognized that the onset of reannihilation in the matter dominated epoch has a strong dependence on the temperature ratio and the mediator mass.
Note that in the parameter region around the reference values in Eq.~\eqref{eq:zrea}, cutoff masses of the order of $10^{8} \, M_{\odot} $ and sizable self-interactions on dwarf galactic scales can be achieved simultaneously. 
Strictly speaking, the simple power-law scaling in Eq.~\eqref{eq:zrea} is only valid for $z_{\text{rea}} \ll z_{\text{eq}}$, where $z_{\text{eq}} \simeq 3400$ is the matter-radiation equality redshift~\cite{Ade:2015xua}, and when the first freeze-out is not significantly affected by Sommerfeld corrections.
We discuss a more general expression for $z_{\text{rea}}$ in Appendix~\ref{sec:zrea} that will later be used in Section~\ref{sec:analysis} to identify the parameter region where reannihilation happens after recombination.

Another region of interest to identify is where reannihilation stops in the radiation dominated era, because here a change in the DM abundance has in general less impact on, e.g., the Hubble expansion rate.
This situation occurs if the saturation temperature $T^{\text{sat}}_{\gamma}$ is higher than the matter-radiation equality temperature $T^{\text{eq}}_{\gamma} = 0.80 \, \text{eV}$.
The saturation temperature $T^{\text{sat}}_{\gamma}$ as a function of the free parameters is derived in Eq.~\eqref{eq:saturationtemperature}.
From this equation it can be read off that the minimum value of $T^{\text{sat}}_{\gamma}$ is given by the minimum $\alpha_{\chi}$ value that can give a resonance.
This occurs when $n = 1$ in Eq.~\eqref{eq:resonancecondition}, and is given by
\be
\alpha_{\chi}^{\text{min}} =\frac{\pi^{2}}{6} \frac{m_{\phi}}{m_{\chi}} \, .
\ee
Inserting this into the result of the saturation temperature [Eq.~\eqref{eq:saturationtemperature}], we find
\bea
T^{\text{sat}}_{\gamma}  &\gtrsim& T_{\gamma}^{\text{sat, min}} = \nonumber \\
&&\hspace{-1.7cm} 0.6 \, \text{eV} \left( \frac{r_{\text{sat}}}{0.36} \right)^{-1} \left( \frac{m_{\phi}}{2 \, \text{GeV}} \right)^{4} \left( \frac{m_{\chi}}{\text{TeV}} \right)^{-7/2} \left( \frac{T^{\text{kd}}_{l}}{\text{MeV}} \right)^{1/2} \,, \label{eq:satbeforemreq}
\eea
and in the case of $\alpha_{\chi} = \alpha_{l} $ the kinetic decoupling temperature in terms of the minimum coupling is given by
\be
T^{\text{kd}}_{l} = 1 \, \text{MeV} \left( \frac{r_{\text{kd}}}{0.5} \right)^{-1/2} \left( \frac{m_{\chi}}{\text{TeV}} \right)^{3/4} \left(\frac{m_{\phi}}{2 \, \text{GeV}} \right)^{1/2} \,.
\ee
Note that the result in Eq.~\eqref{eq:satbeforemreq} is quite general and can be used to estimate the parameter region where one does not expect to have reannihlation below a certain temperature.
It is independent of the physics happening before kinetic decoupling 
and only assumes that the maximum enhancement is given by the $s$-wave unitarity bound and that the saturation temperature is lower than the kinetic decoupling temperature.

Even though we focus on a vector mediator model here, any DM setup where $s$-wave annihilation is Sommerfeld enhanced via a Yukawa potential can lead to an epoch of reannihilation.
Or, more general, any DM model where the total cross section scales as $(\sigma v_{\text{rel}}) \propto  v_{\text{rel}}^{-1-\epsilon}$, with $\epsilon>0$, can lead to an epoch of reannihilation.
This excludes, in particular, $p$-wave annihilation or Coulomb potentials to have the feature of a reannihilation epoch.

\subsection{Numerical methods}
\label{sec:nummeth}
In the previous sections, we established when reannihilation can start and for how long it can last. We now turn to investigate its exact impact on the DM relic density and the Hubble expansion. To track the DM number density [Eq.~\eqref{eq:defnumberdensity}] and the injected energy density evolution during reannihilation, we set up the following coupled differential equations:
\bea
\! \frac{\text{d}Y}{\text{d}x} \! &=& - \frac{s}{\tilde{H} x} \langle \sigma v_{\text{rel}} \rangle_{y} Y^{2} \label{eq:Yevo} \, , \\
\! \frac{\text{d}y}{\text{d}x} \! &=& \! -  \frac{2 \gamma}{\tilde{H} x} \! \left[y \! - \! y^{\text{eq}}\right]\! +\! \frac{s}{\tilde{H} x} y Y \! \left[ \langle \sigma v_{\text{rel}} \rangle_{y} \! \! - \! \langle \sigma v_{\text{rel}} \rangle_{y, 2} \right] \, , \label{eq:yevo} \\
\! \frac{\text{d}Y_{l}}{\text{d}x} \! &=& - \frac{H}{\tilde{H} x} Y_{l} + \frac{s}{\tilde{H}x} \langle \sigma v_{\text{rel}} \rangle_{y} Y^{2} \, , \label{eq:Yprimeevo}
\eea
where we have defined the dimensionless temperatures as
\bea
y &\equiv& \frac{m_{\chi} T_{\chi}}{s^{2/3}} \, \\
y^{\text{eq}} &\equiv& \frac{m_{\chi} T_{l} }{s^{2/3}} = \frac{m_{\chi} r T_{\gamma}}{s^{2/3}} \, .
\eea
The energy density of the injected dark radiation, given by  
\be
Y_{l} \equiv \frac{\rho_{l}}{m_{\chi} s} \, ,
\ee
is fully included in the Hubble expansion rate 
\be
H^{2} = \frac{8 \pi G}{3} \left[ \rho_{\gamma} + \rho_{\nu} + \rho_{b}  +  \rho_{\text{dark}} + \rho_{\Lambda} \right] \,,
\label{eq:fullhubble}
\ee
where the total dark sector energy density is given by
\be
\rho_{\text{dark}} = 2 m_{\chi} s (Y + Y_{l})\,, \label{eq:rhoDM}
\ee
with the factor of 2 originating from the sum of DM particle and anti-particle contributions.
$\tilde{H}$ is defined as
\be
\tilde{H}= \frac{H}{1+ \frac{1}{3} \frac{T}{g_{s}}\frac{ d g_{s} }{d T} } \, ,
\ee
where the evolution of the SM's entropy degrees of freedom $g_s$ and the effective number of relativistic degrees of freedom $g_\text{eff}$ are taken from Ref.~\cite{Drees:2015exa}.

Equations~\eqref{eq:Yevo} and \eqref{eq:yevo} can be derived from the Boltzmann equation in the limit of non-relativistic DM particles
\be
  \label{diff_boltzmann}
  m_{\chi} \left(\partial_{t} - H \mathbf{p} \cdot \nabla_{\mathbf{p}} \right) f_{\chi} = C^{\text{non-rel}}_{\chi \chi \rightarrow \phi \phi, l l}[f_{\chi}]+ C^{\text{non-rel}}_{\chi l \leftrightarrow \chi l}[f_{\chi}, f_{l}^{\text{eq}}]\,,
\ee
by taking the zeroth
\be
  n_{\chi}
=  g \int \frac{\text{d}^{3} p}{(2 \pi)^{3}} f_{\chi} \, ,
\ee
and the second moment with respect to momentum
\be
\label{eq:dm_temp}
T_{\chi} 
= \frac{g}{3n_{\chi}} \int \frac{\text{d}^{3} p}{(2 \pi)^{3}} \frac{p^{2}}{m_{\chi}} f_{\chi} \, ,
\ee
respectively.
$g$ is the DM internal spin degrees of freedom.
Kinetic decoupling from dark radiation is taken into account by the first term in Eq.~\eqref{eq:yevo} and the impact of annihilation on the DM temperature by the last term in the same equation. Equations~\eqref{eq:Yevo} and \eqref{eq:yevo} were derived for the first time in Ref.~\cite{vandenAarssen:2012ag} and can also be obtained by taking the non-relativistic limit of the more general equations as fully derived in Ref.~\cite{Binder:2017rgn}.
The equations of the latter work include relativistic corrections and also the production of DM, where both Eqs.~\eqref{eq:Yevo} and \eqref{eq:yevo} get correction terms.
In this work, for late kinetic decoupling, it is evident that both corrections can be neglected.
Due to different conventions, the momentum transfer rate $\gamma$ is here defined to be a factor of 2 smaller than in Ref.~\cite{Binder:2017rgn}.

In this work we include for the first time the evolution of the dark radiation governed by Eq.~\eqref{eq:Yprimeevo} and the impact of reannihilation on the Hubble expansion rate as in Eq.~\eqref{eq:fullhubble}. Note that both the direct production of $l$ and the instantaneous decay of the produced vector mediators $\phi$ into fermions $l$ are included in the equations via the \emph{total} averaged cross section $\langle \sigma v_{\text{rel}} \rangle_{y}$.
The first term on the right hand side of Eq.~\eqref{eq:Yprimeevo} captures redshifting of the injected dark radiation, while the second term covers that all DM annihilations instantaneously transform non-relativistic DM particles into dark radiation. 

In the rest of this work we will for simplicity assume that the two couplings $g_{\chi}$ and $g_{l}$ of the vector mediator model, as given in Eq.~\eqref{eq:lagrvec}, are equal.
Order one deviations from this assumption do not influence our analysis, since most of the quantities, like kinetic decoupling temperature, have a minor dependence on $g_{l}$.
Furthermore, in some part of the parameter space bound state formation processes might be relevant~\cite{Petraki:2015hla, Petraki:2016cnz, An:2016gad, Cirelli:2016rnw, Baldes:2017gzw}, but it is beyond the scope of this work to investigate it further.

In order to be able to evaluate the phase-space averaged cross sections, $ \langle \sigma v_{\text{rel}} \rangle$ and $\langle \sigma v_{\text{rel}} \rangle_{2}$  defined as
\bea
 \langle \sigma v_{\text{rel}} \rangle
 &\equiv&
 \frac{g^{2}}{n_{\chi}^{2}} \int 
 \frac{d^{3}p\, d^{3}\tilde p}{(2\pi)^{6}}
 (\sigma  v_{\text{rel}} ) f_{\chi}(p) f_{\chi}(\tilde{p}) \, , \\
\label{therm_av2}
 \langle \sigma v_{\text{rel}} \rangle_{2}
 &\equiv&
 \frac{g^{2}}{T n_{\chi}^{2}} \! \int \!
 \frac{d^{3}p\, d^{3}\tilde p}{(2\pi)^{6}}
 \frac{p^{2}}{3 m_{\chi}} 
 (\sigma  v_{\text{rel}} ) f_{\chi}(p) f_{\chi}(\tilde{p}) \, ,
\eea
one has to make an assumption on the form of the DM phase-space distribution.
In the limit of a larger self-scattering rate than the annihilation rate the following form is motivated:
\be
\label{MB_ansatz}
f_{\chi} = \frac{n_{\chi} (T)}{n_{\chi, \text{eq}}(T_{\chi})}\exp \left(- \frac{m_{\chi} + p^{2} / (2 m_{\chi})}{T_{\chi}} \right) \bigg|_{T_{\chi} = y s^{2/3} / m_{\chi}} \, ,
\ee
where the $T_{\chi}$ evolution is governed via Eq.~\eqref{eq:yevo}.
This ansatz leads to the final form of $ \langle \sigma v_{\text{rel}} \rangle_{y}$ given in Eq.~\eqref{eq:sigmavaveraged} and for simplifying the momentum square weighted annihilation cross section $\langle \sigma v_{\text{rel}} \rangle_{y, 2}$ we refer to the result presented in Ref.~\cite{vandenAarssen:2012ag}.
Let us point out that in the DM temperature evolution equation [Eq.~\eqref{eq:yevo}] the two averaged cross sections appear as the difference $[\langle \sigma v_{\text{rel}} \rangle_{y} - \langle \sigma v_{\text{rel}} \rangle_{y, 2}]$.
For Sommerfeld enhanced cross sections, this difference is always positive since $\langle \sigma v_{\text{rel}} \rangle_{y,2}$ has more integral support at higher momenta, where the annihilation cross section is smaller.
If Sommerfeld enhanced annihilation is still significant we therefore expect that $y$ should increase (DM self-heating) after kinetic decoupling \cite{vandenAarssen:2012ag}.
In Fig.~\ref{fig:PSdist}, we see that this is indeed the case.
Without reannihilation, $y$ would otherwise remain almost constant after the kinetic decoupling ended just above $x \sim 10^6$.
\begin{figure*}
  \includegraphics[width=1.05\columnwidth]{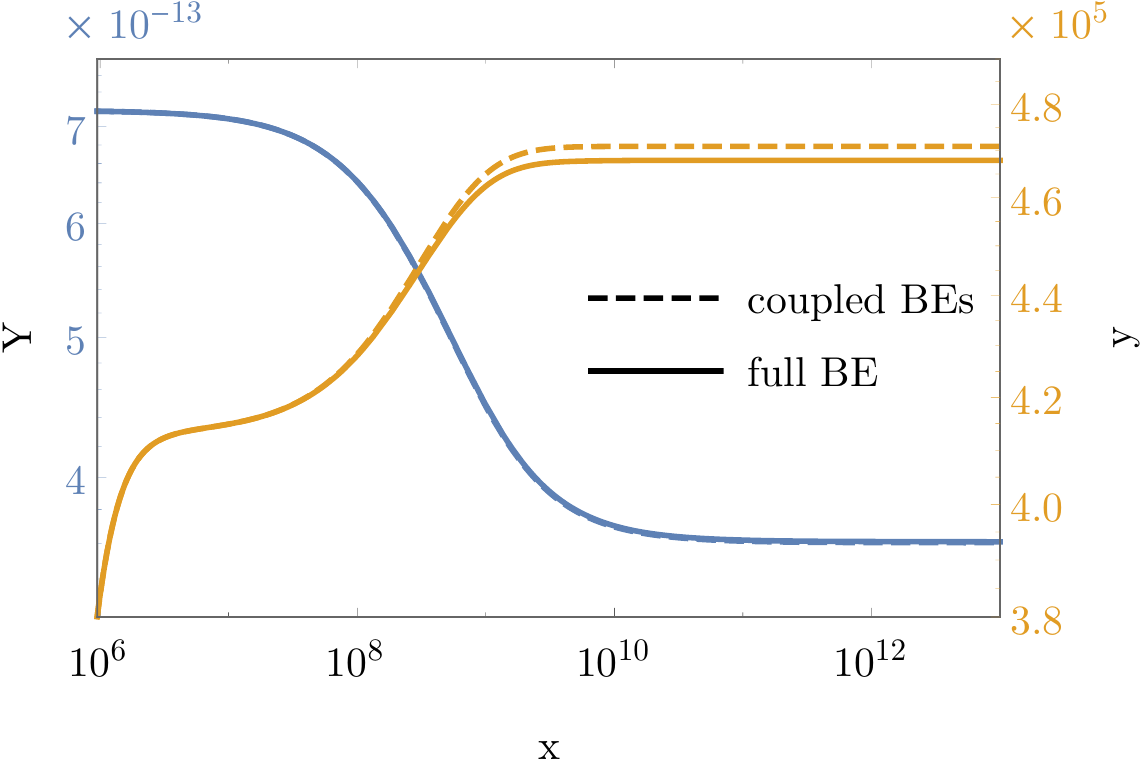}
  \hfill
  \includegraphics[width=0.95\columnwidth]{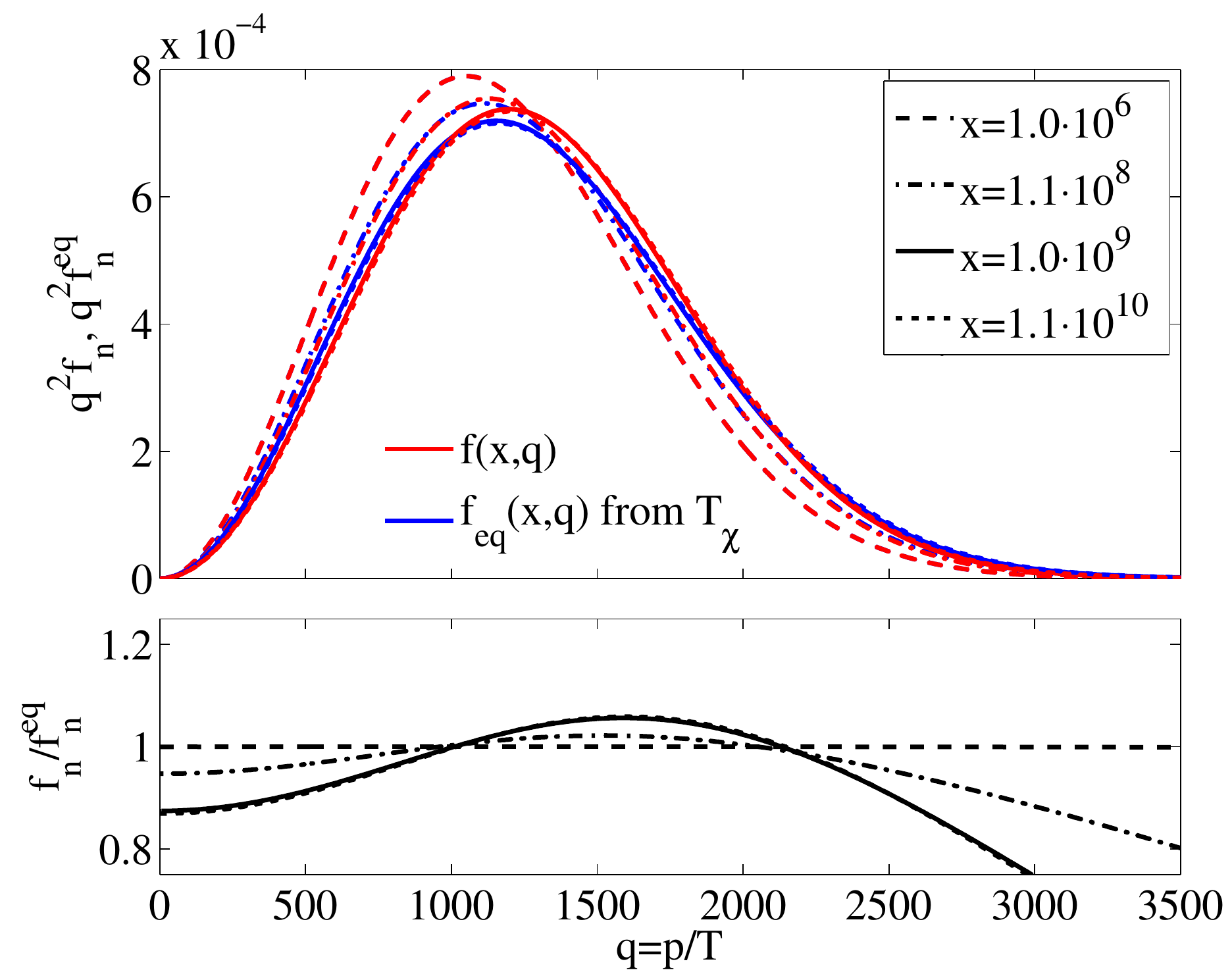}
  \caption{Time evolution of the DM abundance $Y \equiv n_{\chi} / s$, its temperature $y \equiv  \frac{m_{\chi}}{s^{2/3}} T_{\chi}$ and its phase-space density $f(q)$ with $q = p / T_{\gamma}$.
  {\it Left panel:} The evolution of $Y$ (blue) and $y$ (yellow) in the case of strongly self-interacting DM (dotted lines) and in the case of no DM self-interactions (solid).
 {\it Right panel:} Unit normalized phase-space distributions $f_{n} (q)$ from our full numerical solution of the Boltzmann equation (red lines) compared to thermal equilibrium distributions $f^{\text{eq}}_{n} (q)$ with the same ``temperature'' $T_{\chi}$ (blue lines).
  The phase-space distributions are shown at four different $x \simeq $ $10^{6}$ (solid), $10^{8}$  (dashed), $10^{9}$ (dot-dashed) and $10^{10}$ (dotted).
  The bottom panel shows the ratio $f_{n} (q) / f_{n}^{\text{eq}} (q)$.
  The DM model is $m_{\chi} = 600$\,GeV, $m_{\phi} \simeq 1$\,GeV and $\alpha_{\chi}$ chosen such that the relic abundance retains the observed DM abundance after the reannihilation period.
  Both plot styles are chosen to resemble those in Ref.~\cite{Binder:2017rgn}.
   }
    \label{fig:PSdist}
\end{figure*}

It was argued in Refs.~\cite{Feng:2010zp, vandenAarssen:2012ag} that the self-scattering rate can potentially drop below the annihilation rate at the time of reannihilation.
In this case it is possible that the ansatz in Eq.~\eqref{MB_ansatz} is not justified and the momentum moment approach might differ from an exact solution of the full Boltzmann equation.
In the following, however, we confirm for the first time that the momentum moment approach describes \emph{remarkably well} our reannihilation process of Sommerfeld enhanced annihilation, \emph{even in the limit of zero self-scattering}.

Only in the rest of this section, to compare the momentum moment approach in Eqs.~\eqref{eq:Yevo}-\eqref{eq:yevo} to a full phase-space density solution of the Boltzmann equation, we set for simplicity $r=1$, the number of relativistic degrees of freedom to be constant and neglect the impact of reannihilation on the Hubble expansion rate.
We then follow the approach of Ref.~\cite{Binder:2017rgn}, using the dimensionless coordinates
\bea
x(t, p) &\equiv& \frac{m_{\chi}}{T_{\gamma}} \, , \\ 
q(t, p) &\equiv& \frac{p}{T_{\gamma}} \, ,
\eea
to rewrite Eq.~\eqref{diff_boltzmann} for the DM phase-space distribution $f_{\chi} (x,q)$ as
\bea
\p_{x} f_{\chi} (x, q)
 =&& - \frac{m_{\chi}^{3}}{H x^{4}}\frac{g}{4\pi^{2}}\int{d\tilde q\,\tilde q^{2}}\, \int{d \cos{\theta}\,}  \,  (\sigma v_{\text{rel}}) f_{\chi} (q) f_{\chi} (\tilde q)
\nn\\
&&+ \frac{\gamma(x)}{H x} \left[x \partial^{2}_{q} + \left(q + \frac{2 x}{q} \right) \partial_{q} + 3 \right] f_{\chi} \, ,
	\label{eq:BEps}
\eea
where $\theta$ is the angle between the annihilating DM particles' co-moving momenta $\bf{q}$ and $\bf{\tilde q}$.
The Fokker-Planck scattering term has an attractor solution, the non-relativistic Maxwell distribution.
This matches the ansatz in Eq.~\eqref{MB_ansatz} for $T_{\chi} =T_{\gamma}$.

By adapting the code developed in Ref.~\cite{Binder:2017rgn} (to become public~\cite{fBE}), we solve Eq.~\eqref{eq:BEps} and compare its solution to that of Eqs.~\eqref{eq:Yevo}-\eqref{eq:yevo}.
The result around the reannihilation period is presented in Fig.~\ref{fig:PSdist} for one example model.
In the left panel, the solid and dashed blue curves show the DM abundance $Y$ from solving Eq.~\eqref{eq:BEps} and Eqs.~\eqref{eq:Yevo}-\eqref{eq:yevo}, respectively.
After a period of reannihilation starting at $x \sim 10^{8}$, where the effective cross section scales as $1/v^{2}$, the reannihilation stops around $x \sim 10^{10}$ when the Sommerfeld enhancement is saturated and $(\sigma v_{\text{rel}})$ is effectively constant.
In this example, the DM abundance depletes further by $50 \, \%$ during the reannihilation process and converges to the  observed CDM relic density.
The difference in $Y$ between the two approaches is less than $1\, \%$ and the blue curves are virtually overlapping.
So, while it is true that the velocity dependent annihilation cross section acts to heat up DM ---  as shown by the rise of the yellow lines of $y$ around $x \sim 10^{8} \text{--}10^{10}$ after DM kinetically decoupled at $x \simeq 2 \times10^{6}$ --- the distortion of $f_{\chi} (q)$ from a thermal shape is not large enough to significantly alter the relic abundance result.
In the right panel of Fig.~\ref{fig:PSdist}, we show the resulting shape of $f_{\chi}(q)$ (red curves) from the full Boltzmann equation, assuming zero DM self-scattering.
If we compare those (red curves) to reference thermal distributions $f_{\chi}^{\text{eq}}(q)$ (blue curves) that have the same $T_{\chi}$, we see that there is a distortion at the $10\,\%$ level from a thermal equilibrium distributions for $q\lesssim 2500$.
However, this has little effect on the relic abundance because during most of the reannihilation period the effective cross section is close to saturation and varies little with $q$.
The fact that $f_{\chi}$ falls below the corresponding thermal distribution $f_{\chi}^{\text{eq}}$ at larger $q$ does not have any practical implications --- as the number density in the high momentum tail is negligible.
The reason for this $f_{\chi} / f_{\chi}^{\text{eq}}$ suppression is a spurious effect from annihilation at low momenta.
The alteration of $f_{\chi}$ at low momenta leads to a best-fit thermal distribution $f_{\chi}^{\text{eq}}$ with a higher temperature, which in turn is a distribution that has a tail of more large momentum particles.
In the following, we will only investigate small changes in $Y$ and can therefore safely use our system of coupled Eqs.~\eqref{eq:Yevo}--\eqref{eq:Yprimeevo} and \eqref{eq:fullhubble}, which assumes a thermal shape of $f_{\chi}^{\text{eq}}$.

\section{Parameter Scan}
\label{sec:analysis}
Reannihilation leads to the fact that multiple values of $\alpha_{\chi}$ can give the observed DM abundance for fixed model parameters $m_{\chi}$ and $m_{\phi}$.
An initial DM overabundance from the first freeze-out due to lower $\alpha_{\chi}$ values can be compensated by a second period of annihilation.
More precisely this is possible if the three free parameters%
\footnote{We remind the reader that we fix $\alpha_{\chi}=\alpha_{l}$ and $r_{\text{BBN}}=0.5$ with a temperature dependence of $r$ as in Eq.~(\ref{eq:revolution}).}
combine to be close to the parametric resonance condition in Eq.~(\ref{eq:resonancecondition}).
To see this explicitly, we show in Fig.~\ref{fig:alphadeg} the relic abundance as a function of the DM coupling $\alpha_{\chi}$ for fixed values of $m_{\chi}$ and $m_{\phi}$.
The dashed curve shows the relic abundance relative to the correct value if one ignores reannihilation and computes numerically the evolution of the number density in the standard approach~\cite{Gondolo:1990dk}.
\begin{figure}[t]
\centering
\includegraphics[width=0.99\columnwidth]{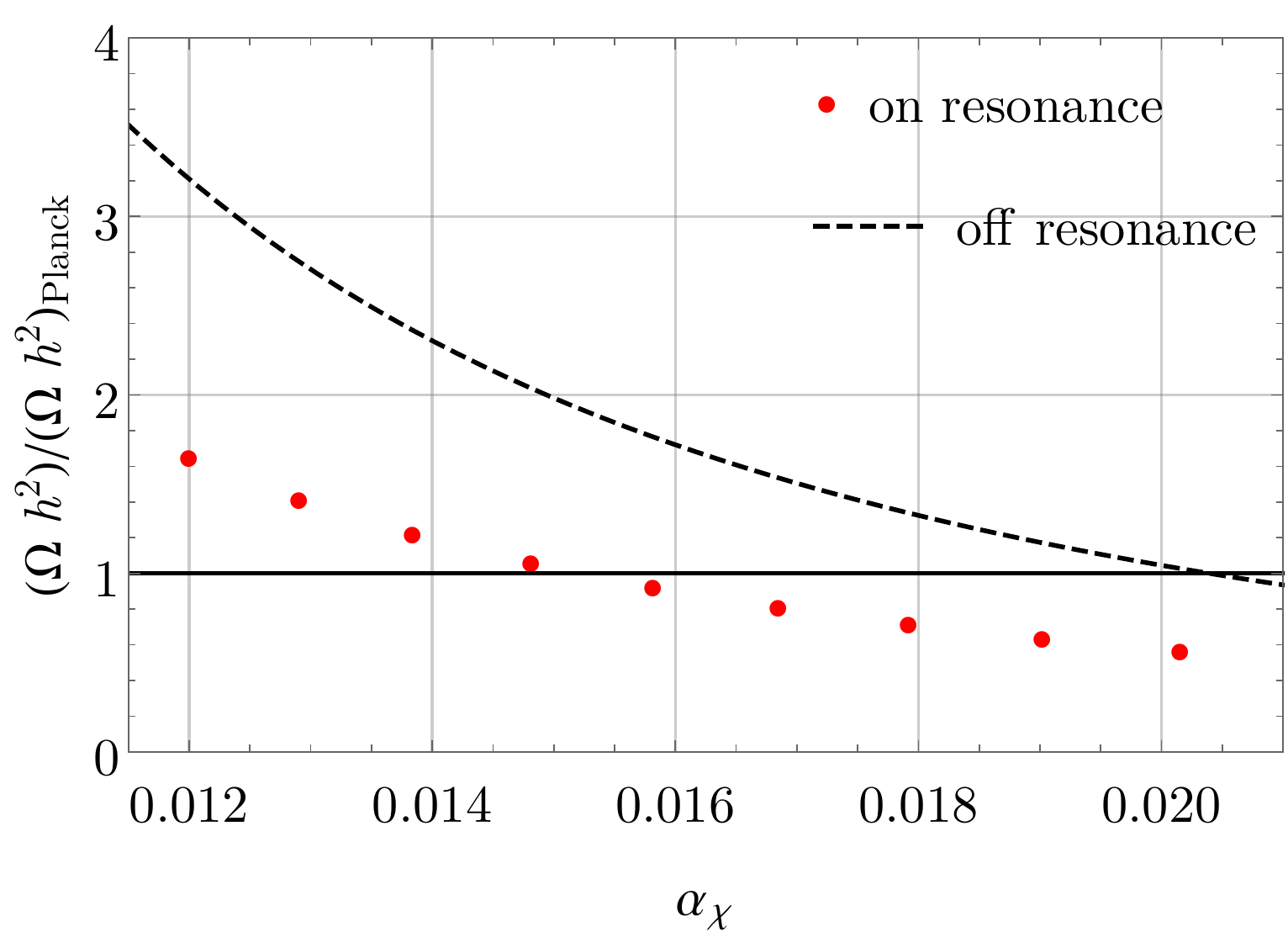}
\caption{Relic abundance ratio shown vs.\ the coupling $\alpha_{\chi}$ for fixed $m_{\chi}=1$\,TeV and $m_{\phi}=10$\,MeV.
Dashed black curve is the result for only taking the standard thermal freeze-out into account (labelled as off resonance).
The red dots present points where the parametric resonance condition is exactly fulfilled and reannihilation thus lowers the relic abundance maximally.
Moving left or right from an exact resonance point by changing $\alpha_{\chi}$ slightly can lead to $(\Omega_{\chi} h^{2})/(\Omega_{c} h^{2})_{\text{Planck}} = 1$ but only for the red points that cross the horizontal black line.
The relic abundance is therefore degenerate in these (almost) on resonance $\alpha_{\chi}$ values.}
\label{fig:alphadeg}
\end{figure}
Clearly, the abundance roughly scales as $\Omega_{\chi}h^{2} \propto \alpha_{\chi}^{-2}$ and there is a unique solution leading to the correct relic abundance at $\alpha_{\chi} \simeq 0.021$ in this example.
It is also demonstrated how the final DM abundance changes by solving Eqs.~\eqref{eq:Yevo}--\eqref{eq:Yprimeevo} numerically for discrete points where the parametric resonance condition is \emph{exactly} fulfilled (red points).
This indicates the \emph{maximal effect} of reannihilation that can be achieved for these $m_{\chi}$ and $m_{\phi}$ values.
The ability to lower $\Omega_\chi$ by reannihilation 
is limited either by the saturation of the Sommerfeld enhancement, or by the finite age of the Universe (where DM halo formation and dark energy domination eventually also halt the reannihilation period).

The fifth red point from the left is the first resonance that can give the correct relic density.
For this point, and the other resonances shown further to the right, there has to be an $\alpha_{\chi}$ in the vicinity of the exact resonance point that reproduces the measured relic density.
In fact there are two $\alpha_{\chi}$ possibilities for each of these resonances since $\Omega_{\chi}h^{2}$ is a smooth function of $\alpha_{\chi}$ that coincides with the off-resonant result between the resonances. Larger values of $\alpha_{\chi}$ than those shown in the figure do not lead to the correct abundance.
To conclude, for given $m_{\chi}$ and $m_{\phi}$ there is a finite number of resonant points that can lead to the correct relic density.
In the example of Fig.~\ref{fig:alphadeg}, in particular, there are five resonances that go below the correct value of the DM abundance and therefore $ 2 \times 5 + 1 = 11$ viable options for $\alpha_{\chi}$.

Having explained above the prescription of counting resonances that result in the correct relic density, we proceed to analyze on-resonance models in a wide parameter range by solving Eqs.~\eqref{eq:Yevo}--\eqref{eq:Yprimeevo} numerically.
We apply the counting prescription to every point on a discrete grid of the order of 0.1 megapixels in the $(m_{\chi}, m_{\phi})$-plane and the result is shown in Fig.~\ref{fig:deg}.
\begin{figure*}
\centering
\includegraphics[width=1.80\columnwidth]{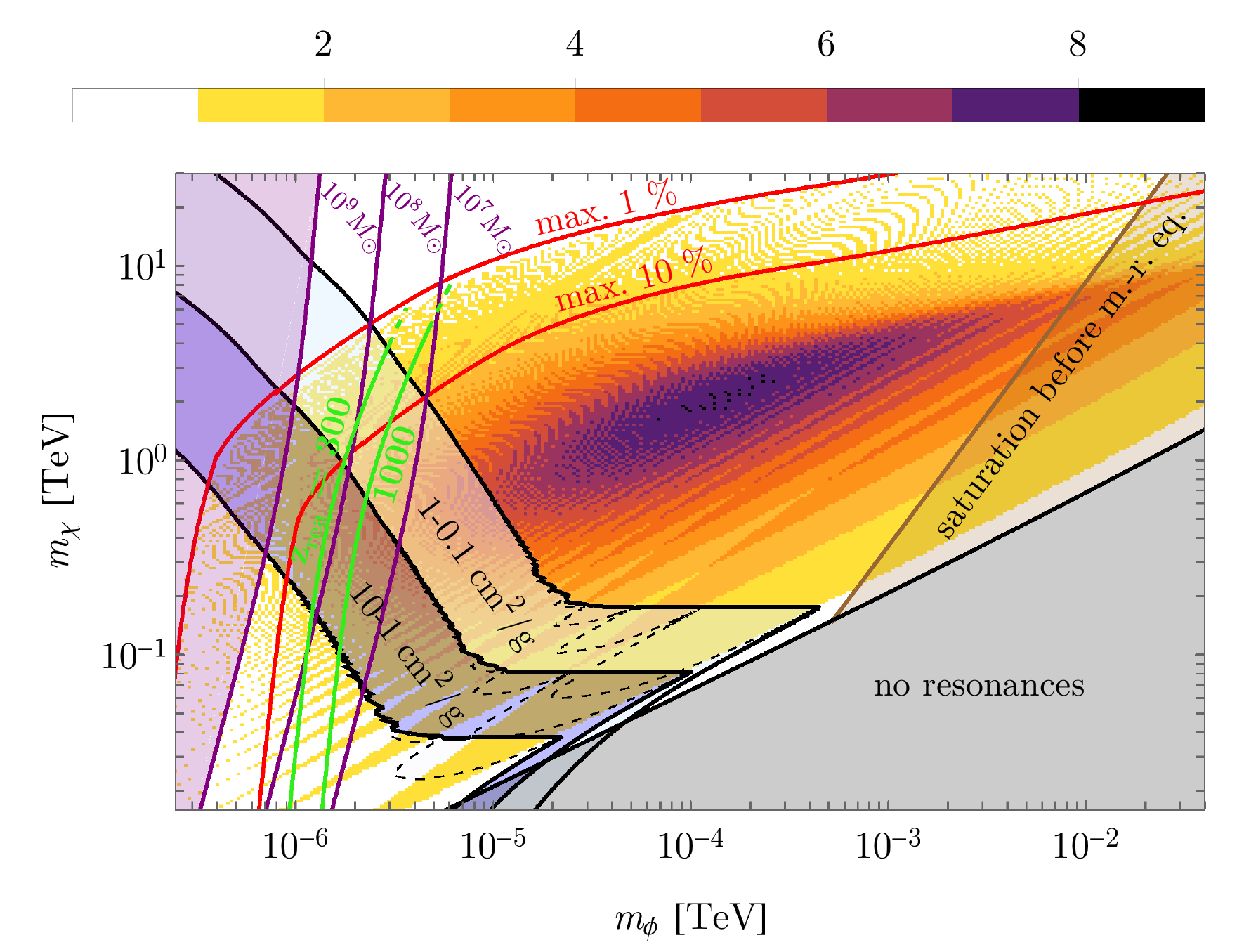}
\caption{Number of Sommerfeld resonances, color-coded as given in the top panel, leading to the correct relic density today and changing the co-moving DM number density by at least $1 \, \%$ during the epoch of reannihilation.
The red solid lines show our analytic estimates (see Appendix~\ref{app:softrea}) of the border where reannihilation can change the relic abundance at most by $1 \, \%$ and $10 \, \%$.
In the shaded grey area in the bottom right part of the figure, no resonances are available leading to the correct relic density.
Brown shaded area represents the estimated region where reannihilation cannot proceed after matter-radiation equality.
Blue and light blue shaded areas cover the parameter space where DM has a sizable self-scattering cross section on dwarf galactic scales: $(\sigma_{T})_{30 \, \text{km/s}}/m_{\chi} \in [0.1,10] \, \text{cm}^{2} \text{g}^{-1}$.
The ``proper'' SIDM region, both in the quantum-resonant and classical self-scattering regime, overlaps with the parameter space where sizable reannihilation can occur.
In the quantum-resonant regime, $\alpha_{\chi}$ is adjusted in the computation of $\sigma_{T}$ such that for given $m_{\chi}$ and $m_{\phi}$ the resonance condition, $\epsilon_{\phi} = 6/(n^{2} \pi^{2})$, is fulfilled for a given integer $n$ (see last subsection of Section~\ref{sec:analysis} for a detailed explanation).
For comparison, the black dashed self-scattering band is for $\alpha_{\chi}$ satisfying the relic density constraint without taking reannihilation or resonances into account.
Cutoff masses of the order of $10^{7}$, $10^{8}$, and $10^{9} \, M_{\odot}$ in the halo-mass function are represented by the purple lines.
In the stripe between the green lines, reannihilation induces the first decrease of the DM co-moving number density by $1\, \%$ between redshifts of $z = 300$ and $z= 1000$ --- while the maximal change in the DM abundance can be read off from the red lines.
In the parameter space where the blue region, the green lines and the purple lines all overlap, SIDM could at the same time alleviate several small-scale structure formation problems and tensions between cosmological parameters derived from CMB and low-redshift astronomical observations (see Section~\ref{sec:cosmogeneral} and Fig.~\ref{fig:handdars}).} 
\label{fig:deg}
\end{figure*}
In our counting algorithm we further require that a resonance should have a sizable impact, i.e., we request the DM relic abundance to change by at least $ 1\, \%$ in order to be counted.
It can be recognized that multiple options of $\alpha_{\chi}$ values exist in a huge parameter space region and in the ``hot spot'' around $m_{\chi} \simeq 2$\,TeV and $m_{\phi} \simeq 100$\,MeV we can have up to $2 \times 8 + 1= 17$ viable $\alpha_{\chi}$ values.

The region of multiple $\alpha_{\chi}$ values is bordered from below by the grey shaded region, where no resonances leading to the correct relic density exist because $\epsilon_{\phi}^{\text{min}} \equiv \frac{m_{\phi}}{\alpha^\text{max}_{\chi} m_{\chi}} > \frac{6}{1^{2} \pi^{2}}$, where $\alpha^{\text{max}}_{\chi}$ represents the maximum (nearly-on-resonant or off-resonant) coupling value leading to the correct relic abundance.
From above and to the left, the region is bounded by the requirement of enabeling at least a $ 1\, \%$ change in the DM relic abundance due to reannihilation; as implemented in the counting algorithm.
In Appendix~\ref{app:softrea}, we provide analytical estimates to explain this ``max  $1\, \%$'' boundary where reannihilation  can not longer significantly change the DM abundance.
The analytical estimates are presented in Fig.~\ref{fig:deg} in terms of the red curves representing the boundaries where reannihilation can maximally change the abundance by $ 1\, \%$ and $ 10\, \%$, respectively.
For points close to the bottom left part of the red lines, reannihilation starts too late in order to change the relic abundance by more than $1\, \%$ and $ 10\, \%$ until today, respectively.
For points close to the red lines where they start to bend for the first time in the left of the plot ($m_\phi\simeq1$\;MeV,\ $m_\chi\simeq$\;TeV), saturation of the Sommerfeld enhancement before today prevents the abundance to change.
In the region where the red curves bend a second time ($m_\phi\simeq10$\;MeV,\ $m_\chi \simeq 10$\;TeV), saturation happens around matter-radiation equality, while in the upper right region saturation takes place in the radiation dominated epoch for counted resonances.
One can clearly see that our estimates match very well the numerical results ($1\, \%$ line).
Let us remark that these analytical estimates can be applied also to other cases, e.g., where $\alpha_{\chi} \neq \alpha_{l}$ or $r_{\text{BBN}} \neq 0.5$ and one does not have to necessarily run a numerical differential equation solver to find these boarders. 

In the brown shaded region reannihilation stops before matter-radiation equality for every resonance. This statement is true even for $\alpha_{\chi}$ values that do not give the correct relic density.
The dashed brown line is the border at which the minimum saturation temperature equals the matter-radiation equality temperature, i.e., $T_{\gamma}^{\text{sat, min}} = 0.80$\,eV  [see Eq.~\eqref{eq:satbeforemreq}]. In Section~\ref{sec:CMB} we will see that the observational consequences of reannihilation are expected to be negligible for models deep inside this brown region.

\subsection*{SIDM region}
It can be recognized in Fig.~\ref{fig:deg} that the SIDM region $(\sigma_{T})_{30 \, \text{km/s}}/m_{\chi} \in [0.1,10] \, \text{cm}^{2} \text{g}^{-1}$ significantly overlaps in the classical and quantum-resonant regime with the region potentially having a sizable reannihilation process. We also show the reannihilation redshift $z_{\text{rea}}$ in terms of the green lines.
In the \emph{most interesting SIDM region}, where also a sizable cutoff mass around $10^{8} \, M_{\odot}$ can be achieved, we conclude from the green lines that reannihilation typically happens in the \emph{matter dominated epoch}.

The computation of the self-scattering cross section $(\sigma_{T})_{30 \, \text{km/s}}$ needs some further consideration in the parameter region where reannihilation happens.
The multiple $\alpha_{\chi}$ values leading to the correct relic abundance would also lead to multiple values of $(\sigma_{T})_{30 \, \text{km/s}}$ for fixed $m_{\phi}$ and $m_{\chi}$.
For the blue self-scattering band in Fig.~\ref{fig:deg} we take the nearly-on-resonant $\alpha_{\chi}$ value that is closest to the off-resonant $\alpha_{\chi}$ leading to the correct relic density.
This is a conservative choice since resonances with lower $n$ would give more sizable reannihilation and thus more often be constrained by, e.g., CMB observations.
In the classical scattering region, this choice has however virtually no impact on the self-scattering band since the resonances are very close to each other and therefore $(\sigma_{T})_{30 \, \text{km/s}}$ does not change significantly when choosing an off-resonant or closest on-resonant value of $\alpha_{\chi}$.
However, in the quantum-resonant regime it makes a significant difference from using an off-resonant value when computing $(\sigma_{T})_{30 \, \text{km/s}}$ as in, e.g., Refs.~\cite{Tulin:2013teo, Tulin:2012wi, Bringmann:2016din, Kahlhoefer:2017umn}, which results in the dashed black curves (where $\alpha_{\chi}$ is uniquely set by the standard relic density constraint, taking no reannihilation into account).
In the Born regime ($\epsilon_{\phi} \gg 1$) we chose $\alpha_{\chi}$ as in the traditional computation since no resonances are available and therefore the coupling is unique.

Let us comment on the choices of fixed $z_{\text{rea}}$ contours given by the green lines in Fig.~\ref{fig:deg}.
Our calculations show that a reannihilation process at $z_{\text{rea}} \simeq 300$ with $5\, \%$ change in the DM abundance starts to saturate between a redshift of $z\sim 30-50$, when most of DM is already confined in virialized halos. Our homogeneous and isotropic treatment of the Boltzmann equation is expected to break down in this non-linear regime due to the increase of DM particle velocities in gravitationally bounded structures.
Therefore we regard $z_{\text{rea}} \simeq 300$ as a lower critical value above which $\sim 5\, \%$ change in the abundance can be achieved.
The reannihilation process starting at $z_{\text{rea}} \simeq 1000$ with $\sim 10\, \%$ change in the abundance saturates much earlier than the time when most of the structures become non-linear and should therefore be safe from this caveat. For redshifts just above $z_{\text{rea}} \simeq 1000$ many CMB quantities might be effected strongly since reannihilation happens around recombination.
A simple approximation of the green lines can be obtained by solving Eq.~\eqref{eq:zrea} for fixed $z_{\text{rea}}$. However, this equation is strictly speaking only valid in the regime where $z_{\text{rea}} \ll z_{\text{eq}}$ and therefore a not good approximation in the case of $z_{\text{rea}} \simeq 1000$. The green lines in Fig.~\ref{fig:deg} are the solution of an improved equation discussed in detail in Appendix~\ref{sec:zrea}.

\section{Cosmological impact}
\label{sec:CMB}
The change in the DM number density and the redshifting of injected dark radiation during reannihilation modifies the expansion rate of the Universe when compared to the $\Lambda$CDM cosmology. Since this process is time dependent, the naive constraints on extra relativistic degrees of freedom $\Delta N_{\text{eff}}$ cannot be applied in general.
Instead, we suggest that the following basic quantities derived from time integration of the modified Hubble expansion rate should not be strongly affected; otherwise reannihilation would hardly reproduce the measured CMB anisotropies or the baryon acoustic oscillation observed in galaxy clustering.

The angular size of the sound horizon $\theta_{*}$ at $z = z_{*}$, where $ z_{*}$ is defined as the redshift where the optical depth $\tau$ equals unity~\cite{Ade:2013zuv}, is a geometrical quantity directly related to the peak positions in the CMB power spectrum and thus precisely measured.
We will work with the value reported by the Planck 2015 (TT+lowP) analysis~\cite{Ade:2015xua}:
\be
100 \theta_{*}= 1.04105 \pm 0.00046 \, \label{eq:planck15theta},
\ee
along with 
\be
z_{*} = 1090.09 \pm 0.42\,.
\ee
From Ref.~\cite{Ade:2013zuv} we have 
\be
100 \theta_{*} = 100 \times r_{s}(z_{*}) / D_{\text{A}} (z_{*}) \label{eq:thstar} \,.
\ee
The sound horizon $r_{s}$ and angular diameter distance $D_{\text{A}}$ are given by
\bea
r_{s} (z) &=& \int^{1/(1+z)}_{0} \frac{da}{a^{2} H \sqrt{3 (1+R)}} \,,\label{eq:rs}\\
D_{\text{A}} (z) &=& \int^{1}_{1/(1+z)} \frac{da}{a^{2} H} \label{eq:da} \,,
\eea
where
\be
R = \frac{3 \rho_{b}}{4 \rho_{\gamma}} = \frac{3 a \, \Omega_{b} h^{2}}{4 \, \Omega_{\gamma} h^{2}} \,,
\label{eq:R}
\ee
and $a$ is the cosmological scale factor.
$r_{s} (z_{*})$ captures the information of the Hubble expansion rate before recombination while $D_{A} (z_{*})$ is sensitive to that between recombination and today.
The definition and further explanation of the introduced quantities can be found in Ref.~\cite{Ade:2013zuv}.
The standard Hubble expansion rate is given by
\be
H^{2}= \frac{8 \pi G}{3} \left[\rho_{\gamma} + \rho_{\nu} + \rho_{c} + \rho_{b} + \rho_{\Lambda} \right] \,.\label{eq:hlcdm}
\ee
In Appendix~\ref{sec:stdhubble}, we provide the details of the cosmological parameters we use to render the above quantities compatible with the Planck 2015 (TT+lowP) measurements~\cite{Ade:2015xua}.
This set of parameters defines our standard Hubble expansion rate of the $\Lambda$CDM cosmology.
When including reannihilation we will replace the standard CDM energy density $\rho_{c}$ with the quantity given in Eq.~\eqref{eq:rhoDM}.
Note that there might exist a compensation between the reannihilation effect and, e.g., the choice of the SM neutrino masses $m_{\nu}$ entering the parametrization of energy density $\rho_{\nu}$ in Eq.~\eqref{eq:hlcdm}.
However, we do not consider this possibility here and fix $m_{\nu}$ as in the Planck 2015 (TT+lowP) analysis~\cite{Ade:2015xua}.
Next we show how the basic quantities given above are sensitive to reannihilation.

\subsection{Reannihilation before recombination}
We here consider reannihilations starting in the radiation dominated epoch and explore the impact on $100 \theta_{*}$. In particular, we investigate the case where the DM abundance is initially overabundant by a few percent and reannihilation leads to the correct observed value.
The evolution of the DM number density and the modified Hubble expansion rate are shown in Fig.~\ref{fig:reabmreq} for such a few scenarios.
It can be seen that the modified Hubble expansion rate starts to increase relative to standard $\Lambda$CDM around the transition from radiation to matter dominated epoch, which is due to the initial overabundance of DM.
It can be recognized that although reannihilation has already saturated around recombination $z_{*}$, the Hubble expansion rate is still modified afterwards.
This can be explained by the gradual redshifting of the injected dark radiation, which delays the return to the standard Hubble expansion rate.

We consider now the impact of the modified expansion rate on $100 \theta_{*}$ by investigating the integrations over $H$ as they appear in Eq.~\eqref{eq:thstar}.
The naive number of standard deviations away from the reported  $100 \theta_{*}$ value in Eq.~\eqref{eq:planck15theta} are calculated and the results as a function of $z_{\text{rea}}$ for a fixed amount of DM depletion are shown in Fig.~\ref{fig:sigma}.
It can be seen that both scenarios presented in Fig.~\ref{fig:reabmreq}, where the DM abundance was initially enhanced by only a few percent, are in strong tension with the value of $100 \theta_{*}$ constrained by the Planck data.
Furthermore, it can be recognized that the angular size of the sound horizon is sensitive even to percentage changes in the DM abundance in the radiation dominated epoch.
However, the deeper in the radiation dominated epoch the reannihilation process takes place the less impact it has on the sound horizon and the more DM would be allowed to annihilate into dark radiation.
This can be simply understood by the fact that changes in the DM abundance in the radiation dominated epoch have no significant impact on the expansion rate as long as the correct abundance is achieved sufficiently before matter-radiation equality. The process of reannihilation \emph{necessarily} takes place in the radiation dominated epoch for parameters in the brown shaded region of Fig.~\ref{fig:deg}. Note that points on the left side of the brown line can still have saturation either before or after matter-radiation equality.
\begin{figure}[t]
\vspace{-0.5cm}
\centering
\includegraphics[width=0.99\columnwidth]{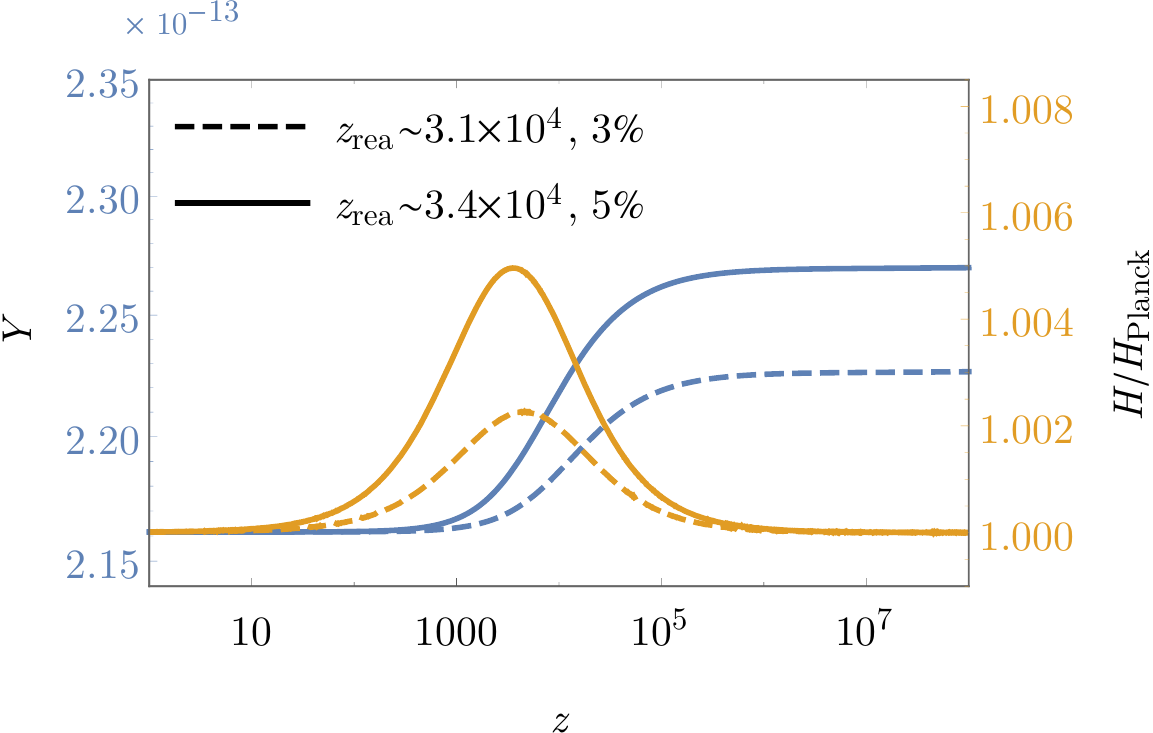}
\caption{Evolution of the DM number density $Y = n_{\chi} / s$ (blue line) and the corresponding  expansion rate $H$ (yellow line) shown as a function of the redshift.
The onset ($1 \, \%$ change in $Y$) of reannihilation for the dashed and solid curves is around $z \simeq 3 \times 10^{4}$ and the DM abundance is initially enhanced by 3 and $5 \, \%$, respectively.
The final relic abundances coincide with $(\Omega_{c} h^{2})_{\text{Planck}} = 0.1197$ and the ratio $H / H_{\text{Planck}}$ therefore reaches 1 at low redshifts.
Both scenarios would be in strong tension with the observed value of $100 \theta_{*}$, see Fig.~\ref{fig:sigma}.}
\label{fig:reabmreq}
\end{figure}
\begin{figure}[t]
\centering
\includegraphics[width=0.79\columnwidth]{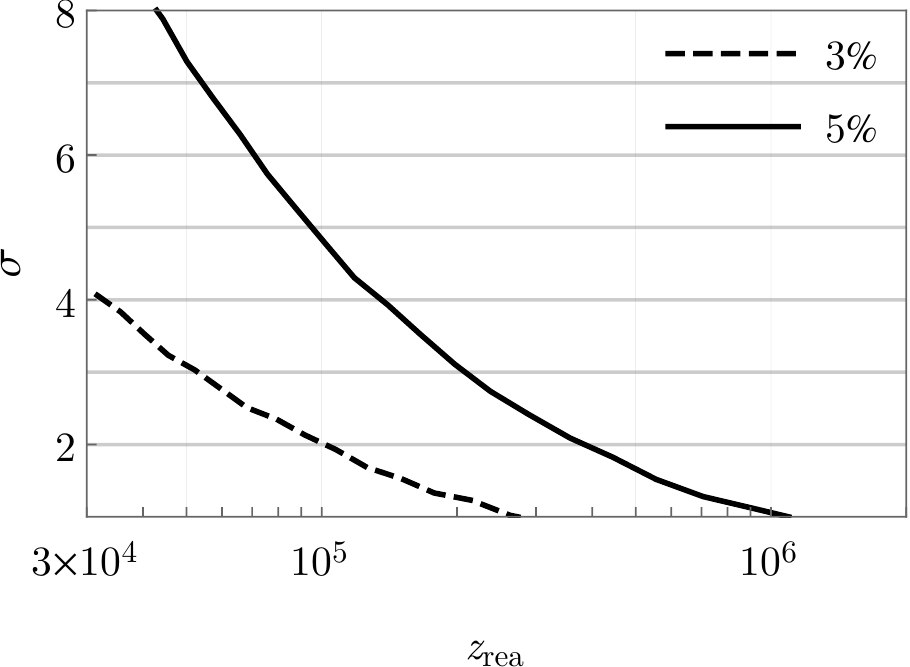}
\caption{Number or standard deviation from the $(100 \theta_{*})_{\text{Planck}}$ measurement vs.\ the redshift of reannihilation onset $z_{\text{rea}}$ (i.e.\ when the co-moving DM abundance first change by more than $1\, \%$).
The two curves refer to 3 and $5\, \%$ total change in the relic abundance where the final value reaches $(\Omega_{c} h^{2})_{\text{Planck}}=0.1197$.
}
    \label{fig:sigma}
\end{figure}

To produce Figs.~\ref{fig:reabmreq} and \ref{fig:sigma}, we used $m_{\chi} = 1$\,TeV and varied $m_{\phi}$ around $\mathcal{O}(10)$\,MeV and adjusted $\alpha_{\chi}$ to have desired DM relic abundance --- but the same result would also be found with other DM model parameters that have the same $z_{\text{rea}}$ and amount of DM depletion during the reannihilation process. 
From our background considerations, we therefore expect that a full Boltzmann code analysis of the CMB would lead to tight constraints on the change in the DM abundance in most of the parameter space in Fig.~\ref{fig:deg} and hence lower the viable number of $\alpha_{\chi}$ values.

\subsection{Reannihilation after recombination}
\label{sec:cosmogeneral}
We now turn to explore the impact on cosmology from reannihilation at late times.
The region of interest is now where reannihilation happens after recombination, $z_{\text{rea}} \lesssim z_{*}$, and especially the area between the green lines in Fig.~\ref{fig:deg}.
It is interesting to note that this area has overlap with both the relevant SIDM region of sizable self-scattering and where the DM halo abundance is suppressed below the mass around $10^{8} \, M_{\odot}$.

The main difference compared to the previous section is that we will here impose compatibleness with the basic CMB quantities [Eqs.~\eqref{eq:thstar}--\eqref{eq:da}] constrained by Planck, while at the same time demonstrate that allowed modifications of the Hubble expansion can alleviate tensions between different cosmological measurements within the $\Lambda$CDM model.
Several works have pointed out the so-called $H_{0}$ tension; a discrepancy within the $\Lambda$CDM model between the measured value of the Hubble constant using CMB data~\cite{Ade:2015xua}, $H_{0} = 67.31 \pm 0.96$\,km\,s$^{-1}$\,Mpc$^{-1}$ (68\,\% C.L.),  and local measurements using only low redshift data, $H_{0} = 73.24 \pm 1.74$\,km\,s$^{-1}$\,Mpc$^{-1}$ (68\,\% C.L.)~\cite{Riess:2016jrr}.
Another tension concerns large-scale structure data and the value of the matter fluctuation amplitude on scales of 8$h^{-1}$\,Mpc, $\sigma_{8}$.
This issue is  related to the $H_{0}$ tension, as the Hubble parameter correlates with the matter density $\Omega_{m}$ and $\sigma_{8}$.
Constraints in the $\sigma_{8}$-$\Omega_{m}$ plane have been widely discussed in the literature~\cite{Battye:2014qga, Salvati:2017rsn, Berezhiani:2015yta, Enqvist:2015ara, Lesgourgues:2015wza, Chudaykin:2016yfk, Poulin:2016nat}, since current CMB data provide significantly different constraints than the thermal Sunyaev-Zel'dovich cluster counts~\cite{Ade:2015fva} and galaxy weak lensing results~\cite{MacCrann:2014wfa, Hildebrandt:2016iqg}, which both prefer lower values of $\sigma_8$.

Our approach will be to require the initial DM abundance to coincide, until recombination, with the reported central value of Planck.
This leaves the sound horizon at recombination unaltered, $r_{s} (z_{*}) = r^{\text{Planck}}_{s} (z_{*})$, since it is a distance derived from integrating $H(a)$ from $a=0$ to the redshift of recombination.
Depending on $z_{\text{rea}}$, reannihilation can then lower the DM abundance after recombination and injects energy in the form of dark radiation until the process saturates.
The loss of DM particles and the redshifting of the dark radiation lowers the Hubble expansion rate $H$ at later times when compared to the $\Lambda$CDM setup, which thus modifies $D_{A}$.
On one hand we require that the tightly constrained quantity $100 \theta_{*}$ is not affected, but on the other hand allow some amount of reannihilation to happen.
This can be achieved by increasing the dark energy content $\rho_{\Lambda}$ in Eq.~\eqref{eq:fullhubble}, such that the period of lower $H$ in the matter dominated epoch is compensated by a period of enhanced $H$ in the dark energy dominated epoch.
In practice, we iteratively change $\rho_{\Lambda}$ to find the desirable $H$ evolution such that $100 \theta_{*}$ does not change when reannihilation lowers the DM abundance.

\begin{figure*}
  \includegraphics[width=0.99\columnwidth]{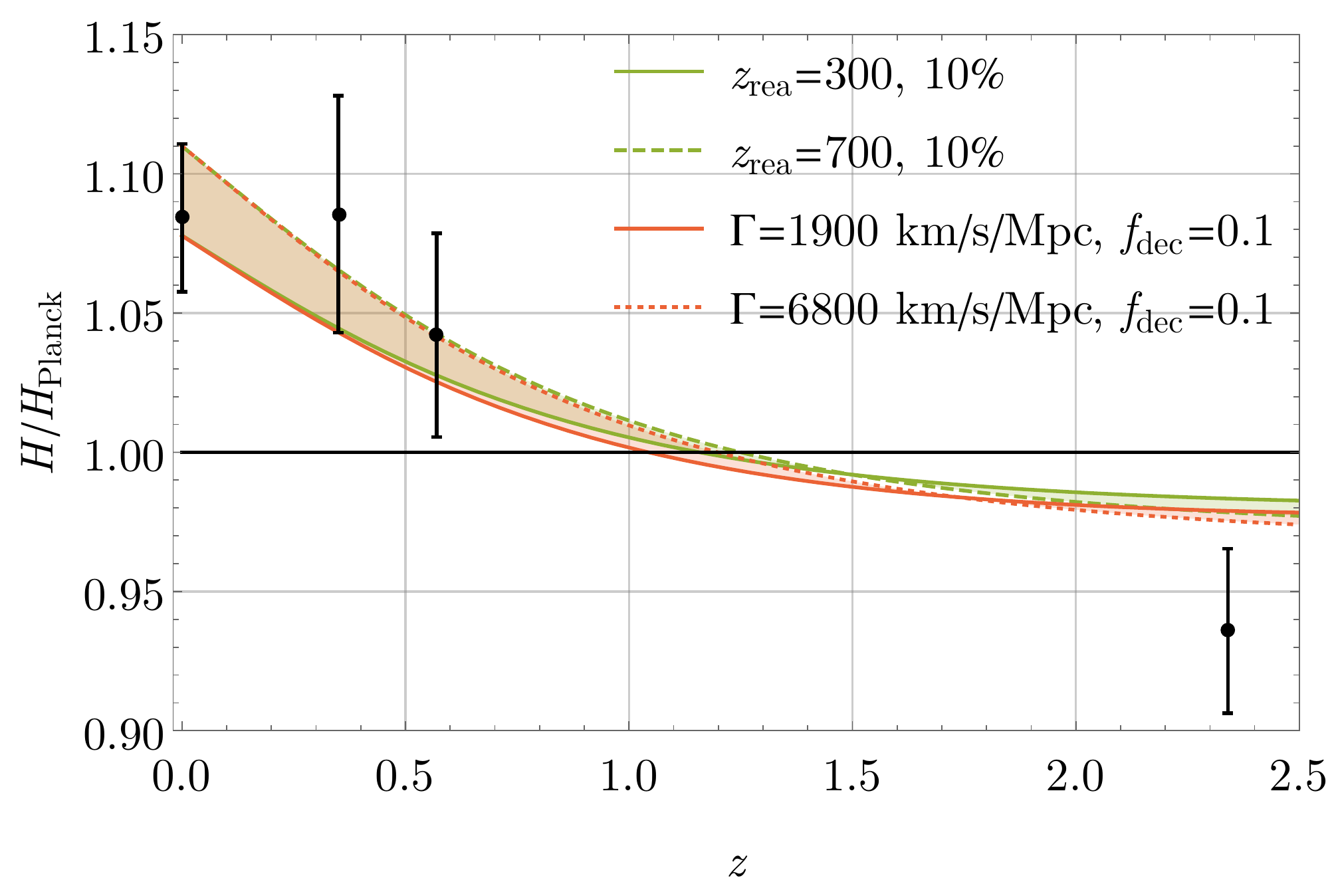}
  \hfill
  \includegraphics[width=0.99\columnwidth]{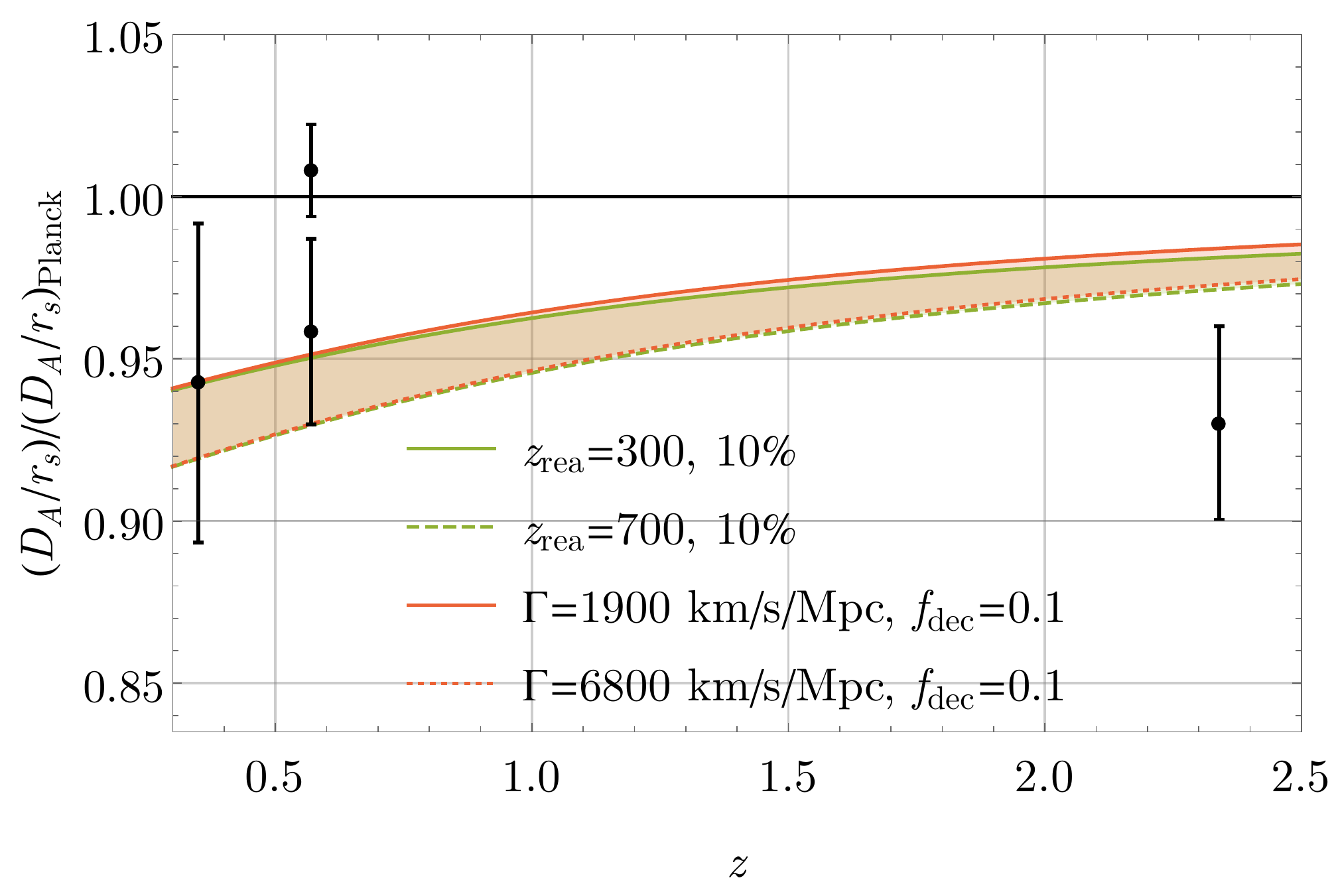}
   \includegraphics[width=0.99\columnwidth]{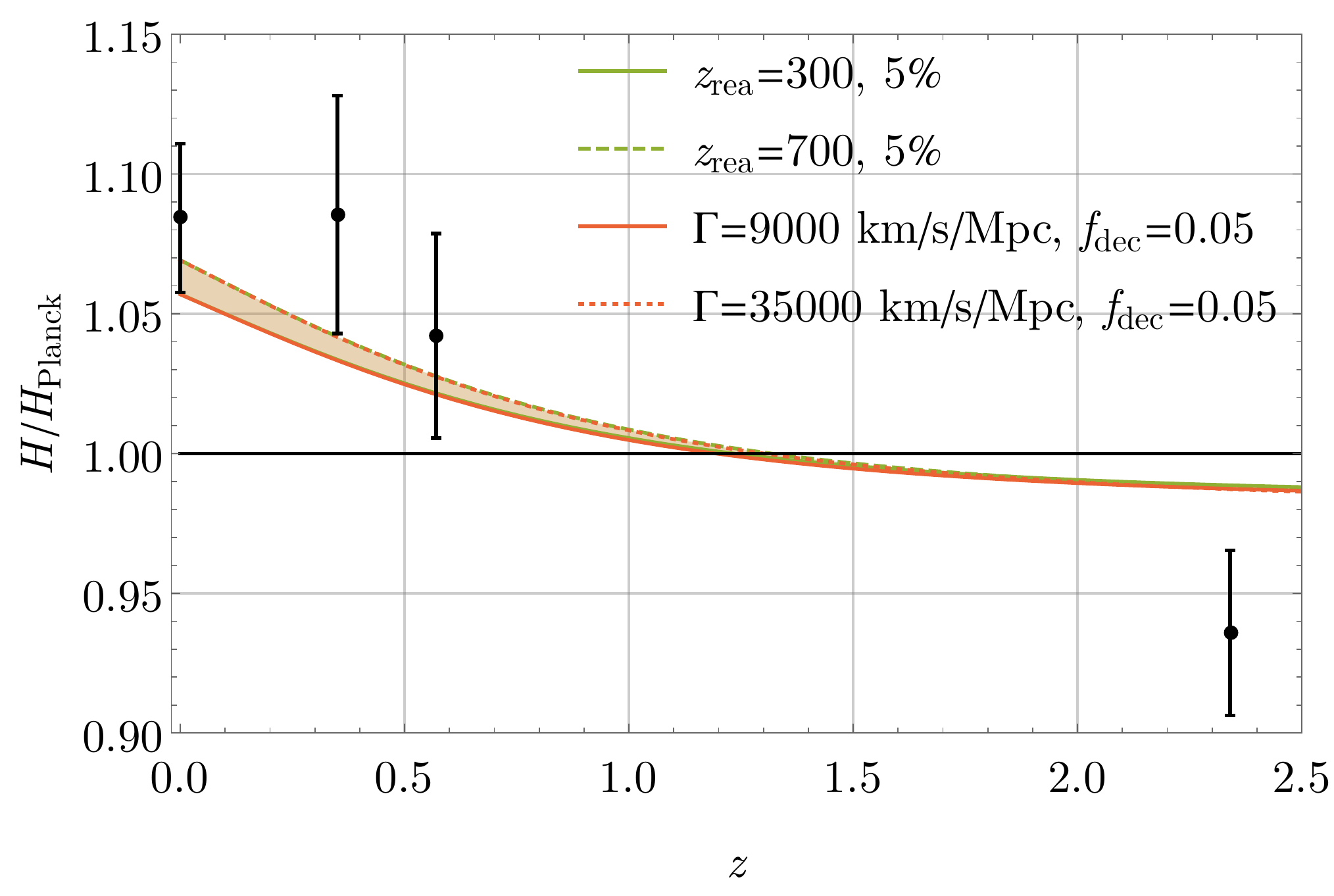}
  \hfill
  \includegraphics[width=0.99\columnwidth]{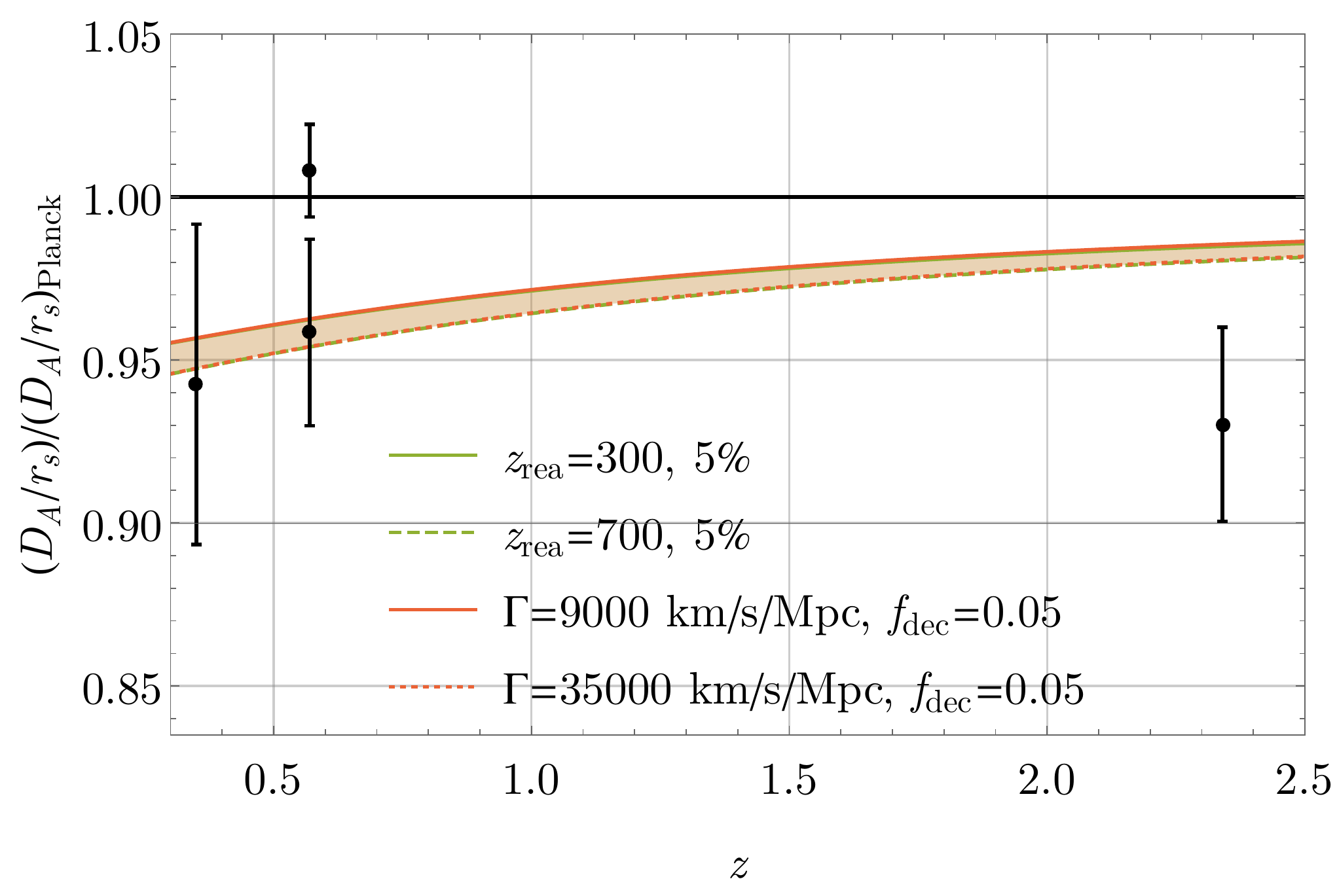}
  \caption{Hubble expansion rate (left) and angular diameter distance (right) ratio  vs.\ redshift for reannihilating (green) and decaying DM (red).
  In both DM scenarios the abundance is changed by $10 \, \%$ (top) and  $5 \, \%$ (bottom) after recombination and the results are almost identical in this redshift interval. Data points are from HST, SDSS and BOSS data \cite{Riess:2016jrr,Chuang:2011fy,Xu:2012fw,Kazin:2013rxa,Anderson:2013zyy,Delubac:2014aqe} and are shown relative to the CMB derived value from Planck data \cite{Ade:2015xua} in the $\Lambda$CDM setup.}
    \label{fig:handdars}
\end{figure*}
The modified expansion rate and the angular diameter distance, computed as explained above, are shown in Fig.~\ref{fig:handdars} together with low-redshift astronomical data: Hubble Space Telescope (HST) at $z=0$~\cite{Riess:2016jrr}, SDSS/BOSS at $z=0.35$~\cite{Chuang:2011fy,Xu:2012fw}, $z=0.57$~\cite{Kazin:2013rxa,Anderson:2013zyy}, and $z=2.34$~\cite{Delubac:2014aqe}.
We demonstrate the modification for $ 5 \, \%$ and $ 10 \, \%$ changes in the DM abundance for $z_{\text{rea}} = 300$ and $700$.
To be in the special SIDM region we have used the parameters $m_{\chi} = 700$\,GeV, $m_{\phi} \in [1.6,2.0]$\,MeV and $\alpha_{\chi}$  tuned to get the  $ 5 \, \%$ and $ 10 \, \%$ changes in the DM abundance, but the same result would be found for every DM model that have the same $z_{\text{rea}}$ and change in the DM abundance (see Fig.~\ref{fig:deg} for further possible options).
In Fig.~\ref{fig:handdars}, one can see that low-redshift data prefers a $6 \text{--}11 \, \%$ larger value of $H_{0}$ than that inferred by the $\Lambda$CDM interpretation of CMB data.
Interestingly, it can be seen that this tension is mitigated by the reannihilation process when changing the DM abundance by $5 \text{--} 10 \%$ after recombination.
The $H / H_{\text{Planck}}$ ratio increases at low redshifts below $z \lesssim 1$ because $\Omega_{\Lambda}$ needs to be larger to keep the highly constrained quantity $100 \theta_{*}$ unchanged.
The reannihilation scenario is also in better agreement with several measurements of the angular diameter distance at low redshifts, while the point reported by Ref.~\cite{Anderson:2013zyy} still favors a pure $\Lambda$CDM cosmology. 

The reduction of $\Omega_{m}$ at low redshifts due to reannihilation leads to a suppressed growth of the matter density perturbations, which might solve the discrepancy in the $\sigma_{8}$-$\Omega_{m}$ plane in $\Lambda$CDM~\cite{Heymans:2013fya, Ade:2015xua}.
The conversion of DM mass density into radiation energy lowers the growth factor since radiation can escape from the gravitational potential and does not contribute to the gravitational growth.
As a consequence, the resultant matter power spectra would be suppressed compared to the $\Lambda$CDM cosmology and thus reannihilation can potentially solves the \mbox{$\sigma_{8}$-$\Omega_{m}$} tension.

The solution of the $\sigma_{8}$-$\Omega_{m}$ tension was discussed for a similar scenario where a part of DM decays into dark radiation after recombination~\cite{Enqvist:2015ara}.
In Fig.~\ref{fig:handdars} we also show our results from a decaying DM scenario, while it was similarly investigated in Ref.~\cite{Berezhiani:2015yta}.
In this setup, the dark sector consists of a DM component of stable $\chi$ particles and mother particles (M) that can decay into effectively massless daughter particles (D).
The energy density evolution of the latter two components can be obtained by solving
\bea
\dot{\rho}_{M} + 3 H \rho_{M}  &=& -\Gamma \, \rho_{M} \,, \\
\dot{\rho}_{D} + 4 H \rho_{D} &=& \Gamma \, \rho_{M} \,,
\eea
numerically, with initial condition $\rho_{M}(t_{i}) = f_{\text{dec}} \rho_{\chi}$ and $\rho_D(t_{i})=0$.
The total dark matter sector's energy density then evolve as
\be
\rho_{\text{dark}}^{\text{decay}} = \rho_{M} + \rho_D + (1 - f_{\text{dec}} )\rho_{\chi} \, . \label{eq:endectotal}
\ee
For comparison, we fix the fraction $f_{\text{dec}}$ of decaying DM (mother particles) with respect to the stable component $\chi$ to $5 \, \%$ and $10 \, \%$, i.e., $f_{\text{dec}} = 0.05$ and $f_{\text{dec}} = 0.1$.
We then match the decay rate $\Gamma$ such that $H_{0}$ coincides with the reannihilation result, while again adjusting the dark energy density to leave the CMB observable $100 \theta_{*}$ unchanged. In Fig.~\ref{fig:handdars} you clearly see that at low redshifts, $z \lesssim 2.5$, the reannihilation and decaying DM models can mimic each other.
They are not distinguishable from these existing astronomical data.

However, let us in the following explain why we believe that these two scenarios impact differently on the evolution of linear perturbations and thus are potentially distinguishable in a CMB power spectrum analysis.
In particular, it was shown in a detailed analysis of Ref.~\cite{Poulin:2016nat} that the CMB observation is still sensitive to decaying DM even long after recombination through the late integrated Sachs-Wolfe effect as it happens in the parameter range as shown in Fig.~\ref{fig:handdars}. The conclusion was that the impact on the CMB power spectrum for the decaying DM scenarios shown in our figures are too strong and essentially ruled out.
However, reannihilation has several different features and the results presented in Ref.~\cite{Poulin:2016nat} for decaying DM cannot be trivially mapped one-to-one to annihilating DM. First of all, it is clear that the energy density of the unstable mother particles $\rho_{M}$ decays exponentially fast in time. In the reannihilation case the DM density changes slower and its duration is longer compared to decaying DM. Therefore, the evolution of the injected dark radiation (or daughter particles) and the modification of the Hubble expansion rate are different. Second, the right hand side of the Boltzmann equation is proportional to DM density squared for reannihilation, while for decaying DM it is instead linear in the density. This might lead to further differences in the evolution of cosmological perturbations. Third, the annihilation process is velocity dependent and one does not expect reannihilation to happen at wavelength modes that have already formed sizable structure. Finally, for large cut-off masses of the order $10^8 M_{\odot}$ it has been found that the reionization history is different compared to $\Lambda$CDM predictions \cite{Lovell:2017eec, Das:2017nub}. We leave a development of a Boltzmann code and a more detailed investigation of these issues to future work.

\section{Summary and Conclusion}
\label{sec:concl}
In this work we have explored the observational imprints of a second period of DM annihilation into dark radiation.
We have shown that such an epoch of reannihilation can arise in DM models where the annihilation cross section is $s$-wave dominated and resonantly Sommerfeld enhanced.
As a concrete realization we have considered a simple model where sizable self-interactions are induced by a light vector mediator, interacting with a dark matter particle and a massless background particle in a fully closed dark sector.
We have extensively analyzed the reannihilation phenomenology of this model and found that this process can change the initial DM number density set by the standard thermal freeze-out, by up to a factor of several in a wide range of the model parameter space.
Furthermore, the onset of reannihilation can range from being deep in the radiation dominated epoch to the beginning of halo formation.
 
In the most interesting parameter region of our considered particle physics model --- where several small-scale structure formation issues can be addressed --- we have shown that the reannihilation process starts during the matter dominated epoch.
Existing CMB data, which is sensitive to even only a few percentage changes in the DM abundance during this epoch, might confirm the existence of such scenarios.
We have interestingly found that in the same parameter region the reannihilation process might reduce the tension between CMB and low-redshift astronomical observations of $H_{0}$ and $\sigma_{8}$ --- although our discussion is limited at the background level.
We have also demonstrated that reannihilation can be used as a clear signature to break the otherwise close degeneracy between scalar and vector mediator realizations of self-interacting dark matter models.
 
The effects on cosmological perturbations, especially on the CMB power spectrum, might be non-trivial even in the cases where reannihilation happens much later than recombination or deep in the radiation dominated epoch.
A dedicated Boltzmann code deserves to be developed to identify the detailed signatures of reannihilation and to clarify how well tensions between CMB and low-redshift astronomical observations can be alleviated.
\vfill

\acknowledgments{ 
\vspace{-0.2cm}
MG and TB thank Torsten Bringmann, Laura Covi, Andrzej Hryczuk, Sebastian Wild, and Hai-bo Yu for reading and commenting on our draft, as well as the participants at SIDM workshop in Copenhagen for stimulating discussions.
AK thanks Ryusuke Jinno and Toyokazu Sekiguchi for useful discussions.
MG and TB have received funding from the European Union's Horizon 2020 research and innovation programme under grant agreement No 690575 and No 674896.
TB gratefully acknowledges financial support from the German Science Foundation (DFG RTG 1493).
The work of AK was supported by IBS under the project code, IBS-R018-D1.
SRS and MW gratefully thank the ITP G\"ottingen for the nice hospitality during the early stage of this project, which is partially based on our Bachelor's theses~\cite{srs:2016, mw:2016}. We all thank Marcel Langenberg for his support with the GWDG computer cluster.
}

\onecolumngrid
\appendix

\section{Theoretical uncertainties in the computation of self-scattering cross sections}
\label{app:ssxs}
In Fig.~\ref{plots} we show a comparison between $\sigma_{T} / m_{\chi}$ in a classical approximation and in the quantum treatment discussed in detail in the Appendix of Ref.~\cite{Kahlhoefer:2017umn}.
From this figure we conclude that there are quantum corrections in both the vector and scalar mediator setups, but for our work they are small enough to be neglected. We note that we see a tendency of an increase of the corrections for higher velocities. This might imply larger corrections on Galactic cluster scales, with $v_{0} \sim 1000 \, \text{km/s}$. To perform a precise calculation on such velocity scales it would require the summation of many more scattering phases $\delta_{\ell}$ which is beyond the scope of this paper. 
\begin{figure*}[!ht]
    \centering
    \subfigure{\includegraphics[width=8.2cm]{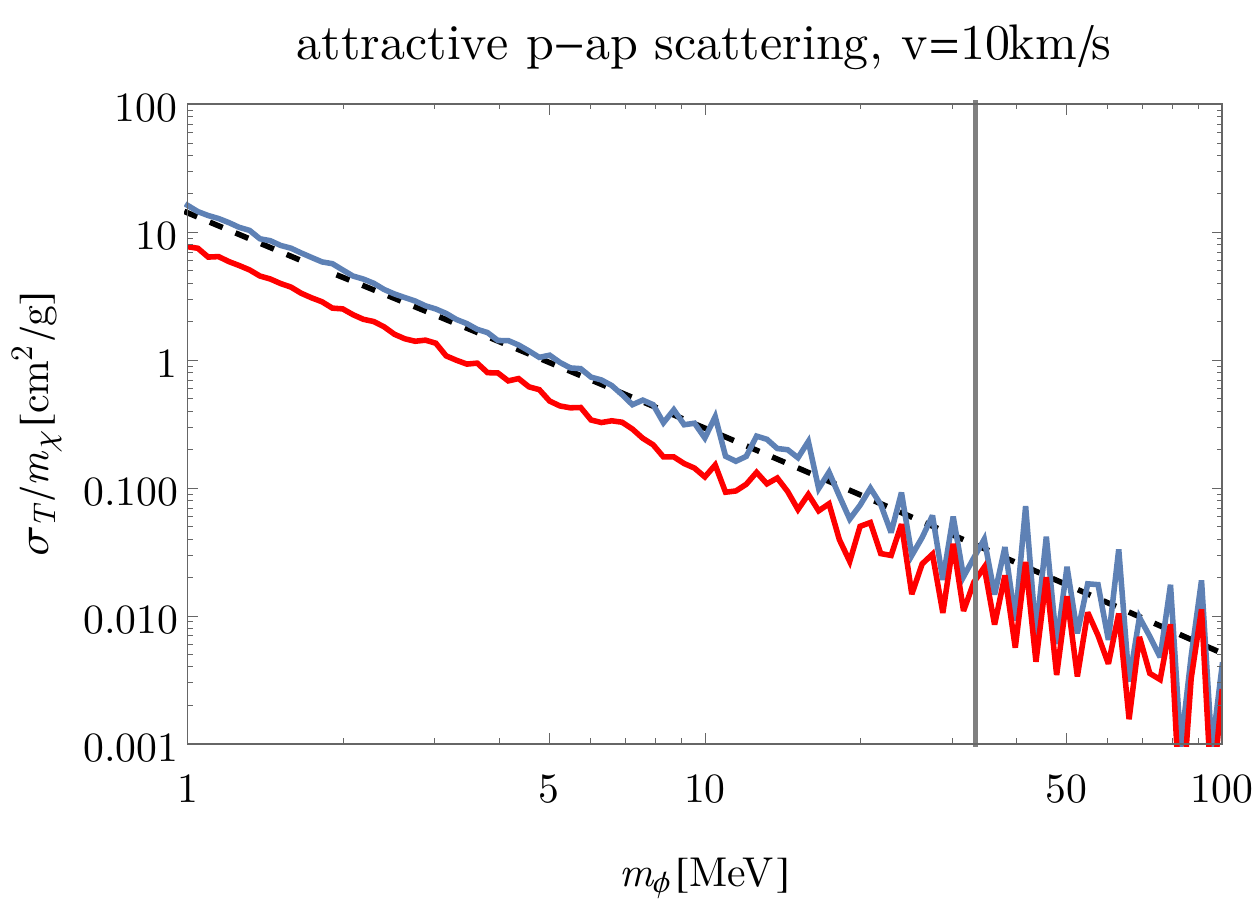}}
    \subfigure{\includegraphics[width=8.2cm]{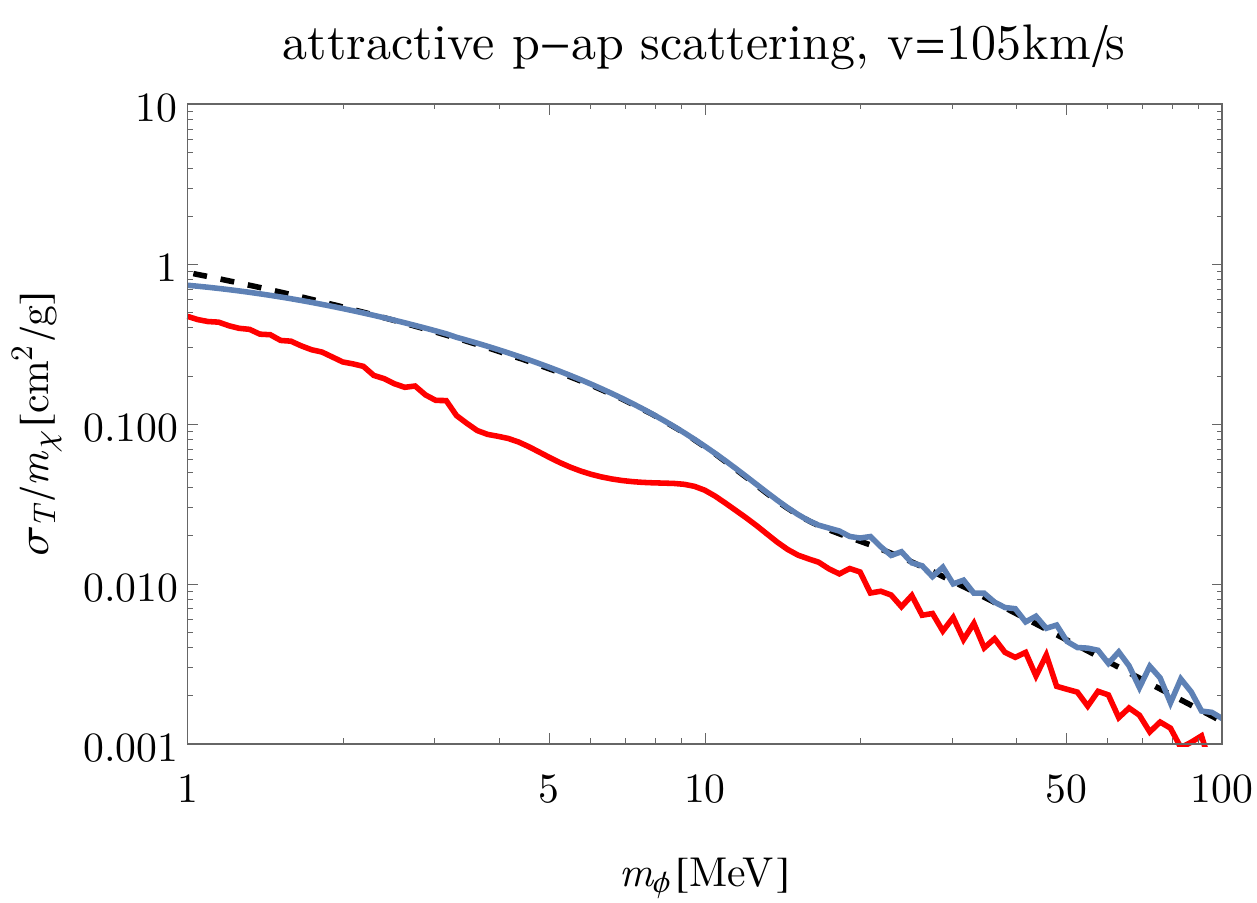}}\\
    \subfigure{\includegraphics[width=8.2cm]{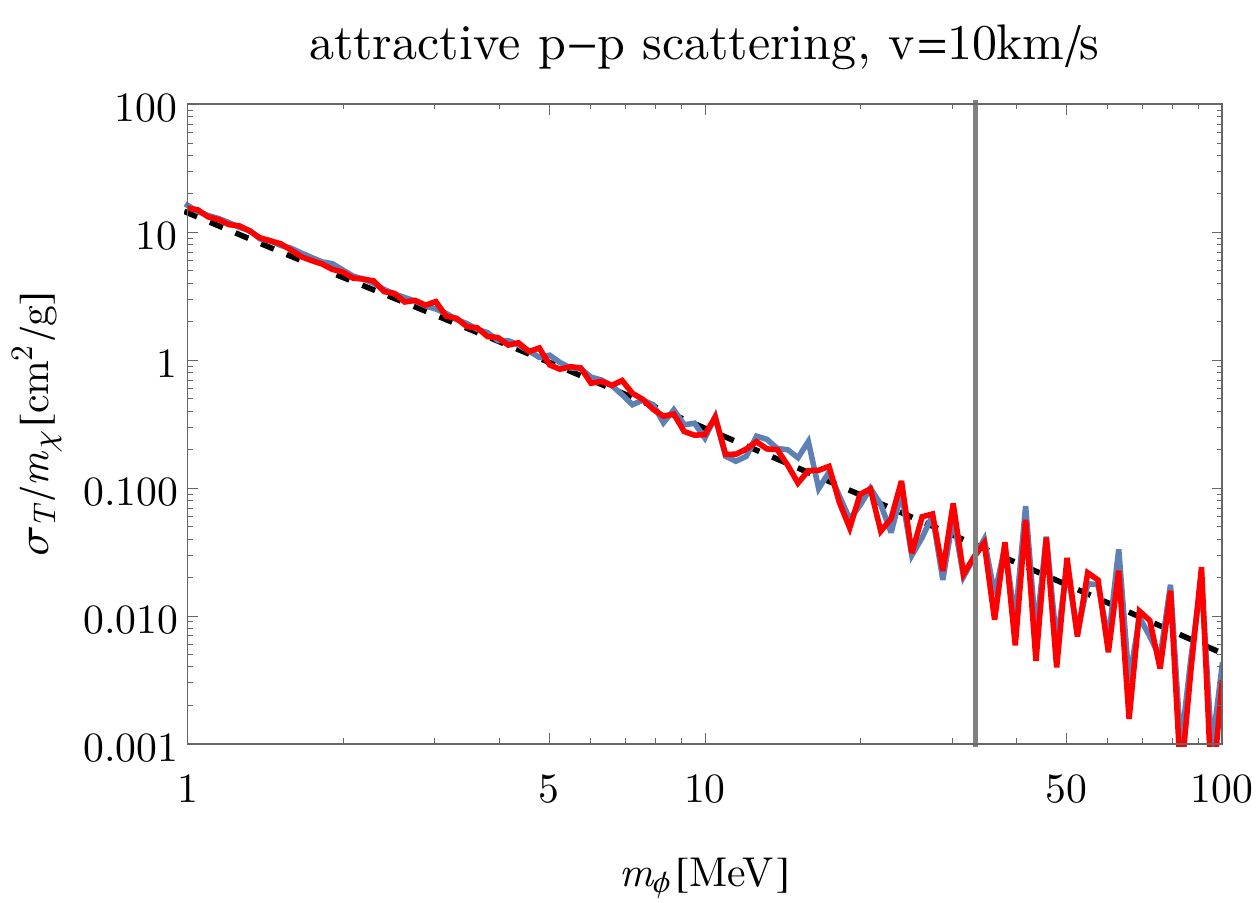}}
    \subfigure{\includegraphics[width=8.2cm]{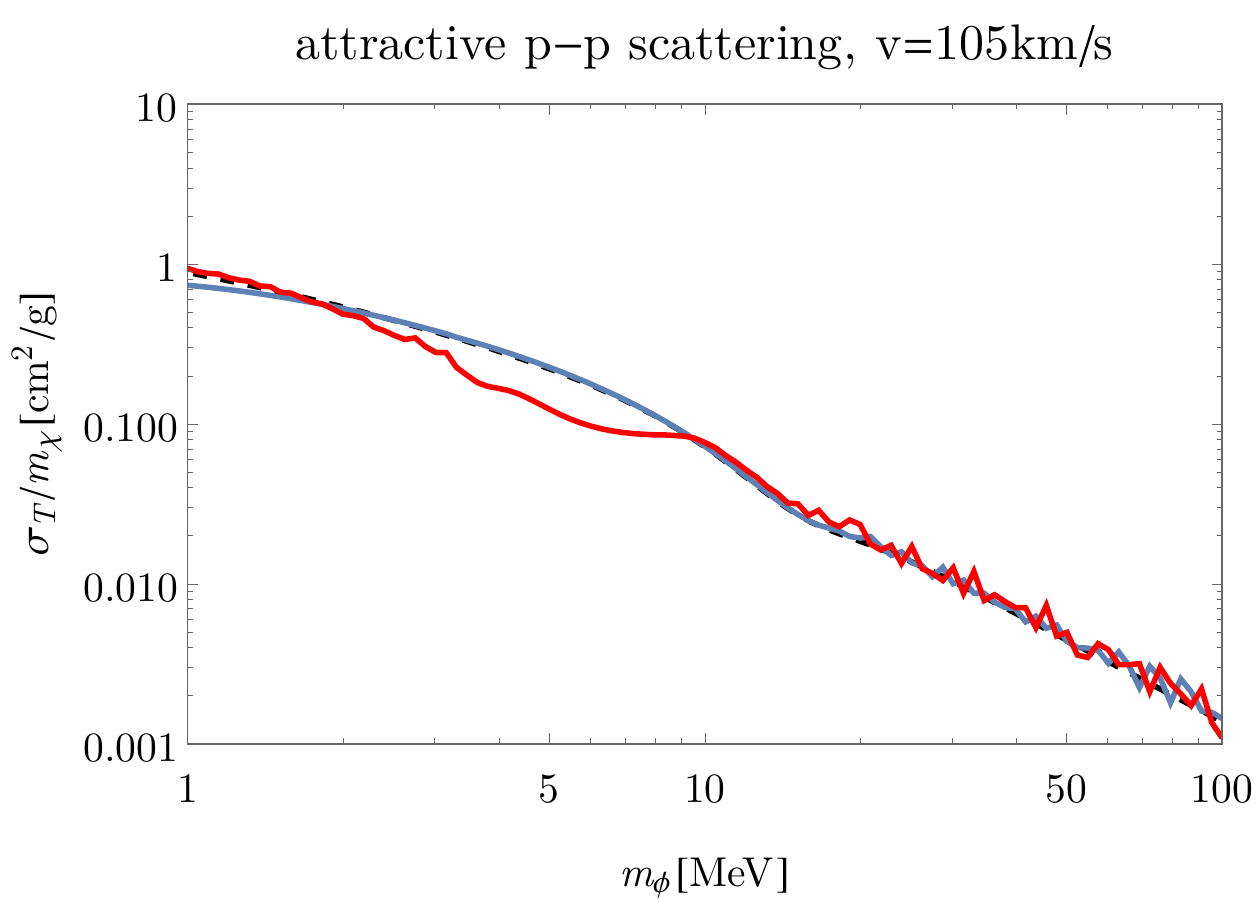}}\\
    \subfigure{\includegraphics[width=8.2cm]{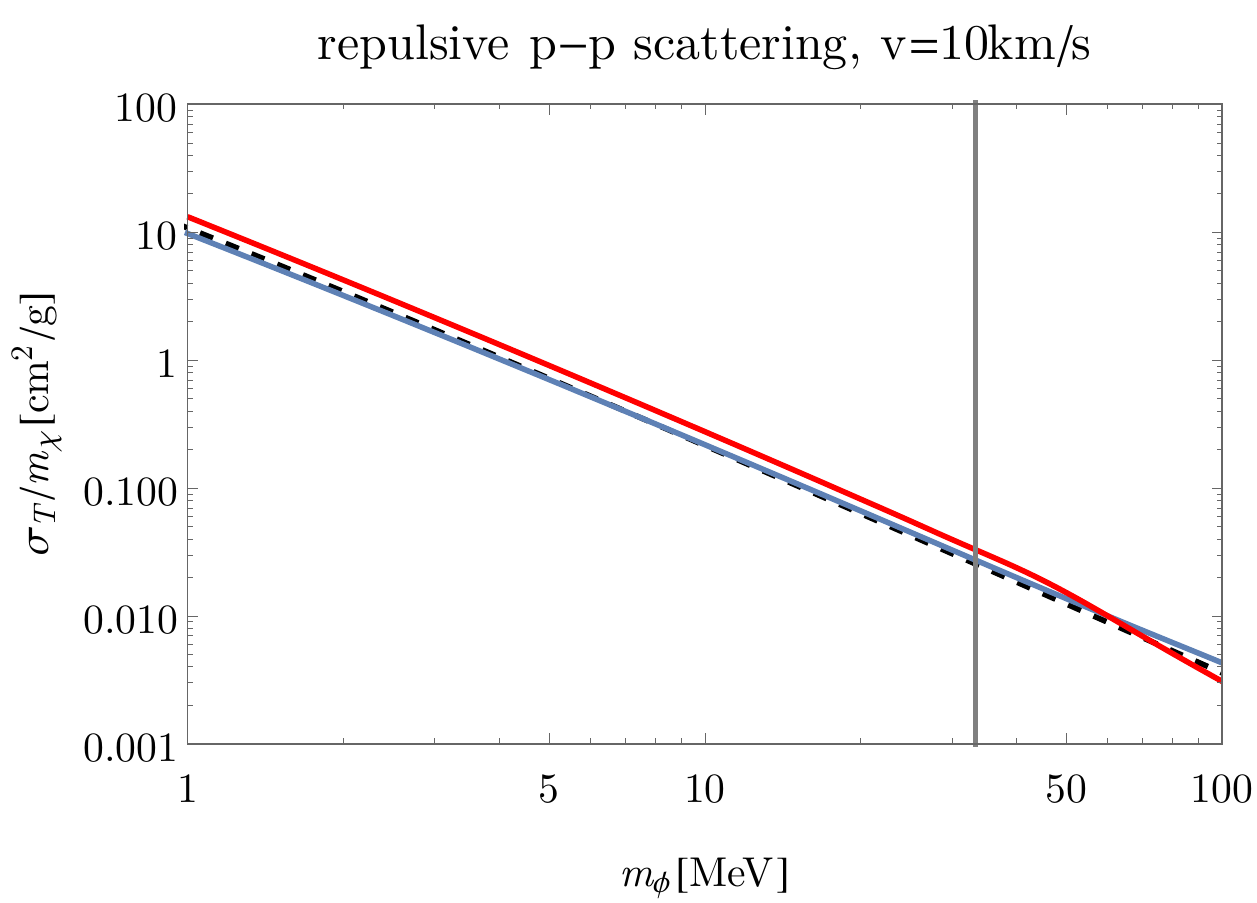}}
    \subfigure{\includegraphics[width=8.2cm]{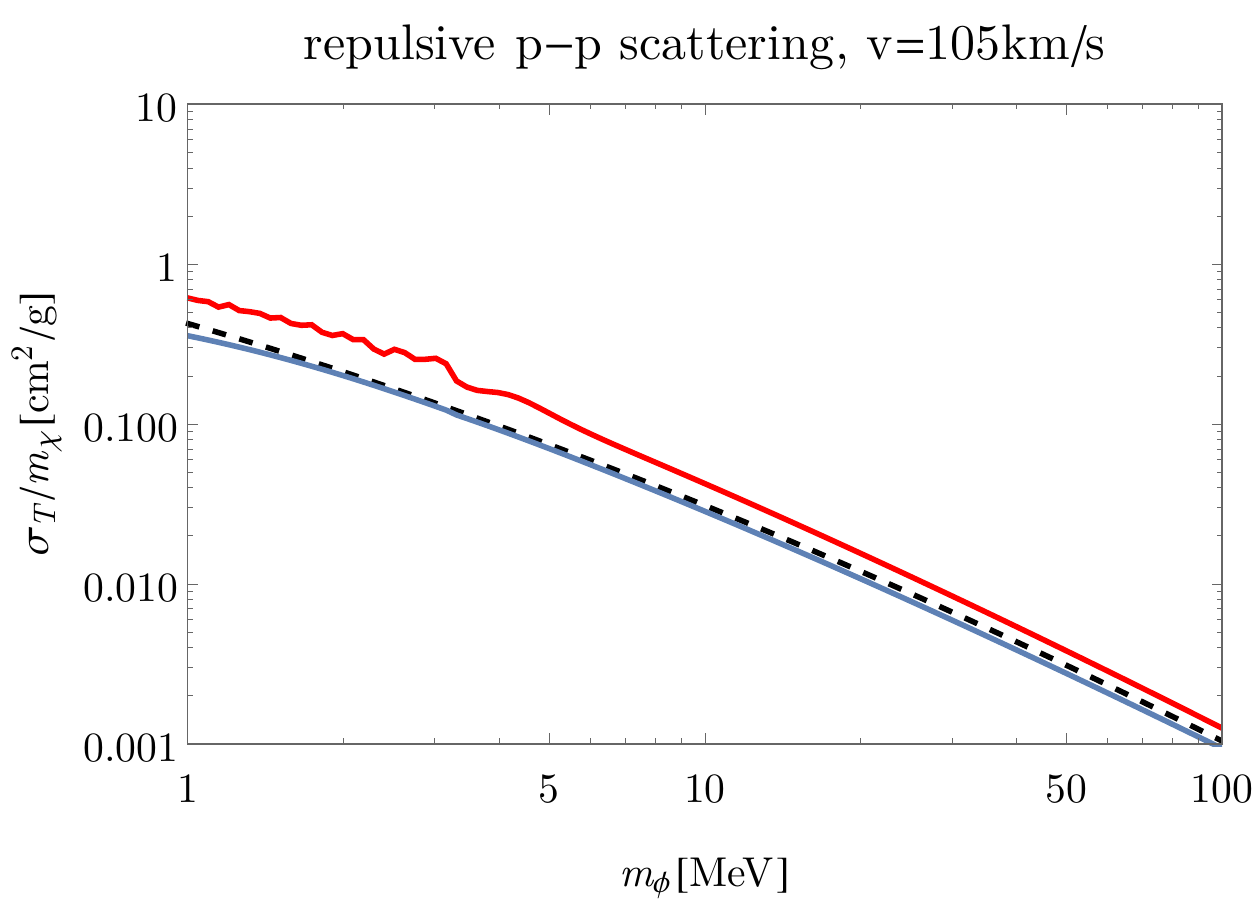}}
    \caption{
Self-scattering transfer cross sections vs.\ mediator mass $m_{\phi}$ from numerical solutions of the Schr\"odinger equation~\cite{Tulin:2013teo,Tulin:2012wi,Kahlhoefer:2017umn} (red and blue lines) compared to the ETHOS~\cite{Cyr-Racine:2015ihg} fitting functions (black dashed lines).  The particle model is fixed to $m_{\chi}=1 \, \text{TeV}$ and $\alpha_{\chi}=0.033$.
 \emph{Left: } is for a relative DM velocity $v_{0}=10 \, \text{km/s}$.
\emph{Right:} $v_{0}=105 \, \text{km/s}$.
\emph{Top:} scattering with an attractive Yukawa potential between particle-antiparticle (p-ap)  (applies to vector and scalar mediators).
\emph{Middle:} attractive particle-particle scattering (scalar mediators).
\emph{Bottom}: repulsive particle-particle scattering (vector mediators).
The numerical solutions includes the computation and summation of phase-shifts $\delta_{\ell}$ up to $\ell = 150$ (left) and $\ell = 225$ (right).
Red curve corresponds to the computation of $\sigma_{T}$ when including quantum statistics and averaging $\text{d} \sigma/\text{d}\Omega$ over $1 - |\cos \theta|$, as suggested in Ref.~\cite{Kahlhoefer:2017umn}. Points to the left of the gray line are in the regime $m_{\chi} v_{0} \gtrsim m_{\phi}$ where the blue and dashed black curve are expected to coincide.
Deviation from the red line indicates the theoretical bias of taking a classical approximation and averaging the scattering amplitude over $1 - \cos \theta$ [see Eq.~\eqref{eq:transfercrx}] instead of $1 - |\cos \theta|$.
}
    \label{plots}
\end{figure*}

\section{Self-consistent description of the Sommerfeld enhancement}
\label{app:selfconsistentS}
It was pointed out in Ref.~\cite{Blum:2016nrz} that close to a resonance it is required to calculate the Sommerfeld enhancement self-consistently in order not to violate the partial wave unitarity limit.
This means that in the derivation of the DM non-relativistic effective theory both the long \emph{and} short range contributions have to be taken into account in the effective potential of the two-body Schr\"odinger equation.
In our scenario the long range part is the Yukawa potential, while the short range contribution consists of the hard annihilation and scattering processes.
The \emph{regulated} formula for the total $s$-wave Sommerfeld enhanced annihilation cross section derived from a self-consistent solution of the Schr\"odinger equation is given for attractive forces by \cite{Blum:2016nrz}
\be
(\sigma v_{\text{rel}})_{\text{ann}} \simeq (\sigma v_{\text{rel}})_{\text{ann}, 0} \times \frac{S (v_{\text{rel}})}{\Biggl|1 + v_{\text{rel}} \left(- \sqrt{ \frac{\mu^{2} \sigma_{\text{sc}, 0}}{4 \pi} - \left( \frac{\mu^{2} (\sigma v_{\text{rel}})_{\text{ann}, 0}}{4 \pi} \right)^{2}} - i \frac{\mu^{2} (\sigma v_{\text{rel}})_{\text{ann}, 0}}{4 \pi} \right) \left( T (v_{\text{rel}}) + i S (v_{\text{rel}}) \right) \Biggl|^{2}} \, . \label{eq:svfull}
\ee
In our work we approximate the Yukawa potential as the Hulth\'en potential for which $S(v_{\text{rel}})$ is given in Eq.~\eqref{eq:sofv} and $T$ takes the form \cite{Blum:2016nrz}
\bea
T (v_{\text{rel}}) &\simeq& - \frac{1}{2 \epsilon_{v}} \left( H (\alpha_{+}) + H (\alpha_{-}) + H (- \alpha_{+}) + H (- \alpha_{-}) - \{p \rightarrow p_{0}\}\right),\\
\alpha_{\pm} &=& i \frac{\epsilon_{v}}{\epsilon_{\phi} \pi^{2} / 6} \pm \sqrt{ \frac{1}{\epsilon_{\phi} \pi^{2} / 6} - \left(\frac{\epsilon_{v}}{\epsilon_{\phi} \pi^{2} / 6}\right)^{2}}\, .
\eea
Here, $H(z)$ is the analytic continuation of the $z$-th Harmonic Number.
For the tree-level annihilation cross section $(\sigma v_{\text{rel}})_{\text{ann}, 0}$ in Eq.~\eqref{eq:svfull} we take the sum over all tree-level channels, 
\be
(\sigma v_{\text{rel}})_{\text{ann}, 0} = \sum_{i} (\sigma v_{\text{rel}})_{0,i} \,,
\ee
as given in Eqs.~\eqref{eq:sv0vecPhi} and \eqref{eq:sv0vecl}.
For the hard self-scattering cross section $\sigma_{\text{sc,0}}$ in Eq.~\eqref{eq:svfull} we take
\be
\sigma_{\text{sc,0}}= \frac{3 \alpha_{\chi}^{2} \pi}{4 m_{\chi}^{2}}\,,\label{eq:sigmascdef}
\ee
which can be obtained from the s-channel diagram of non-relativistic particle-antiparticle scattering.
In our computation of $T(v_{\text{rel}}) $ we drop the matching term $\{p \rightarrow p_{0}\}$, since it is only relevant close to the high energy scale $p_{0}$~\cite{Blum:2016nrz}.

The regulated Hulth\'en potential solution as described above comes with a subtlety discussed in the following.
The short range quantities $(\sigma v_{\text{rel}})_{\text{ann}, 0}$ and $\sigma_{\text{sc}, 0}$ affect the parametric resonance condition slightly when compared to the unregulated solution $S(v)$.
To avoid to repeatedly have to find the precise numerical resonance condition of a regulated solution when studying each single resonance in, e.g., the parameter scan of Fig.~\ref{fig:deg}, we decided throughout this work to approximate $\sigma_{\text{sc}, 0} = \frac{\mu}{4 \pi} (\sigma v_{\text{rel}})_{\text{ann}, 0}^{2}$ such that the square root in the denominator of Eq.~\eqref{eq:svfull} vanishes.
In the right panel of Fig.~\ref{fig:Sv} it is demonstrated that this choice of $\sigma_{\text{sc}, 0}$ only shifts the parametric resonance condition back to the known expression $\epsilon_{\phi} = 6 / (n^{2} \pi^{2})$, however, the hight of the enhancement peak is practically unaffected. By numerical evidence, we have further checked that $(\sigma v_{\text{rel}})_{\text{ann}}$ is modified by at most about 10\,\% for all velocities in all the parameter regions we study. Furthermore, we looked at the numerical solution with the Yukawa potential and demonstrate in Fig.~\ref{fig:Sv} that also in this case only the resonance condition slightly deviates from $\epsilon_{\phi} = 6 / (n^{2} \pi^{2})$.
It can be seen in all cases of the regulated Hulth\'en potential solution that the maximal enhancement respects the unitarity bound of $s$-wave annihilation cross sections, given by (see, e.g., Ref.~\cite{Blum:2016nrz})
\be
\sigma_{\text{max}} = \frac{\pi}{\mu^{2} v_{\text{rel}}^{2}} \, ,
\label{eq:unitaritybound}
\ee
where the reduced mass is here given by $\mu = m_{\chi} /2$.
Naively, the value of the Sommerfeld factor where it saturates can now be obtained from $\sigma_{\text{max}} = (\sigma v_{\text{rel}})_{\text{ann}, 0} S(v_{\text{rel}}) / v_{\text{rel}}$, namely, 
\be
S^{\text{sat}}(v_{\text{rel}}) = \frac{\pi}{\mu^{2} v_{\text{rel}} (\sigma v_{\text{rel}})_{\text{ann}, 0}} \, . \label{eq:smax}
\ee
This expression will be used in Appendix \ref{app:reaestimates} to estimate the saturation velocity of the Sommerfeld factor.

\begin{figure}
\centering{
\includegraphics[width=0.49\columnwidth]{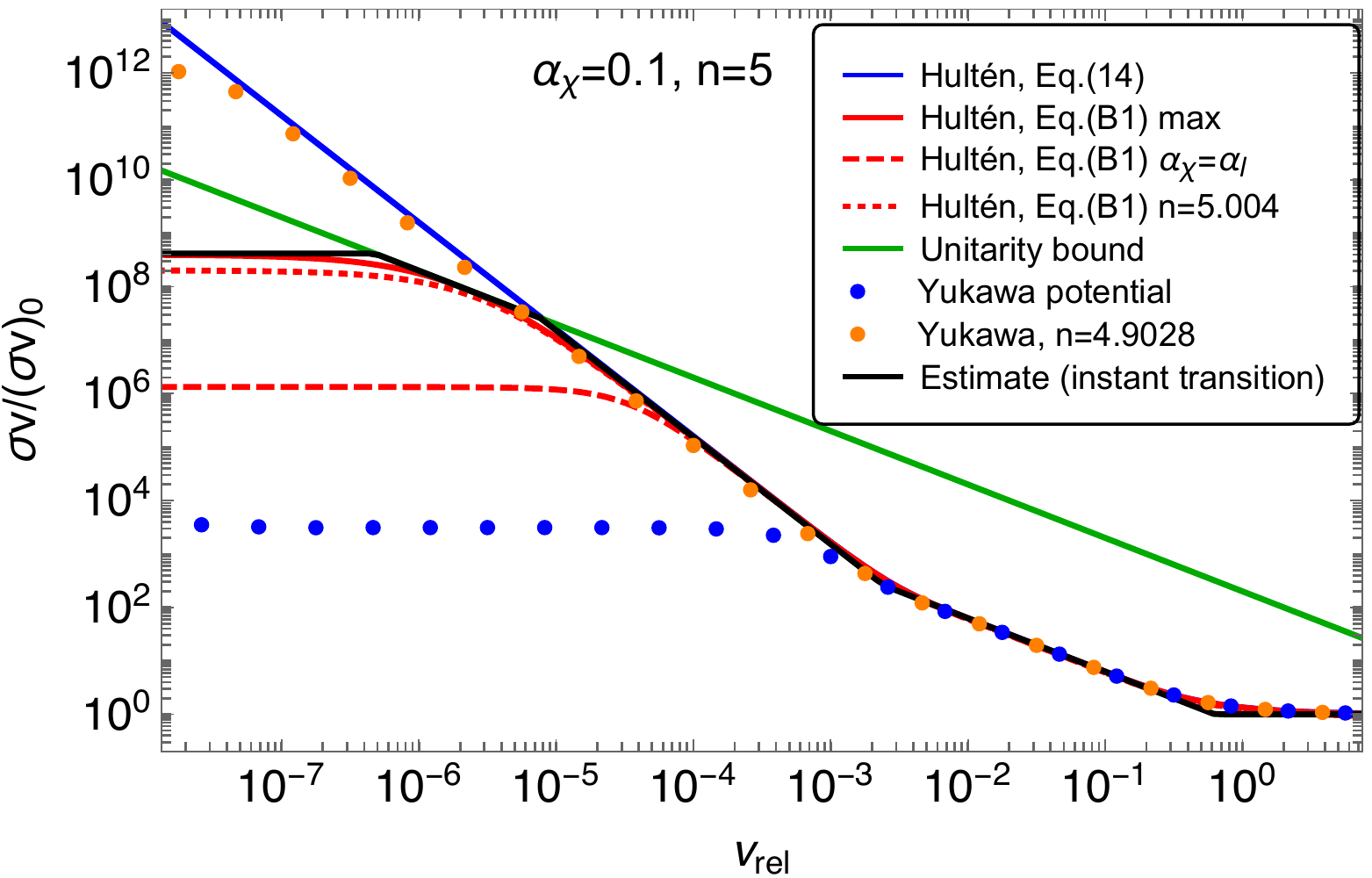}
\includegraphics[width=0.49\columnwidth]{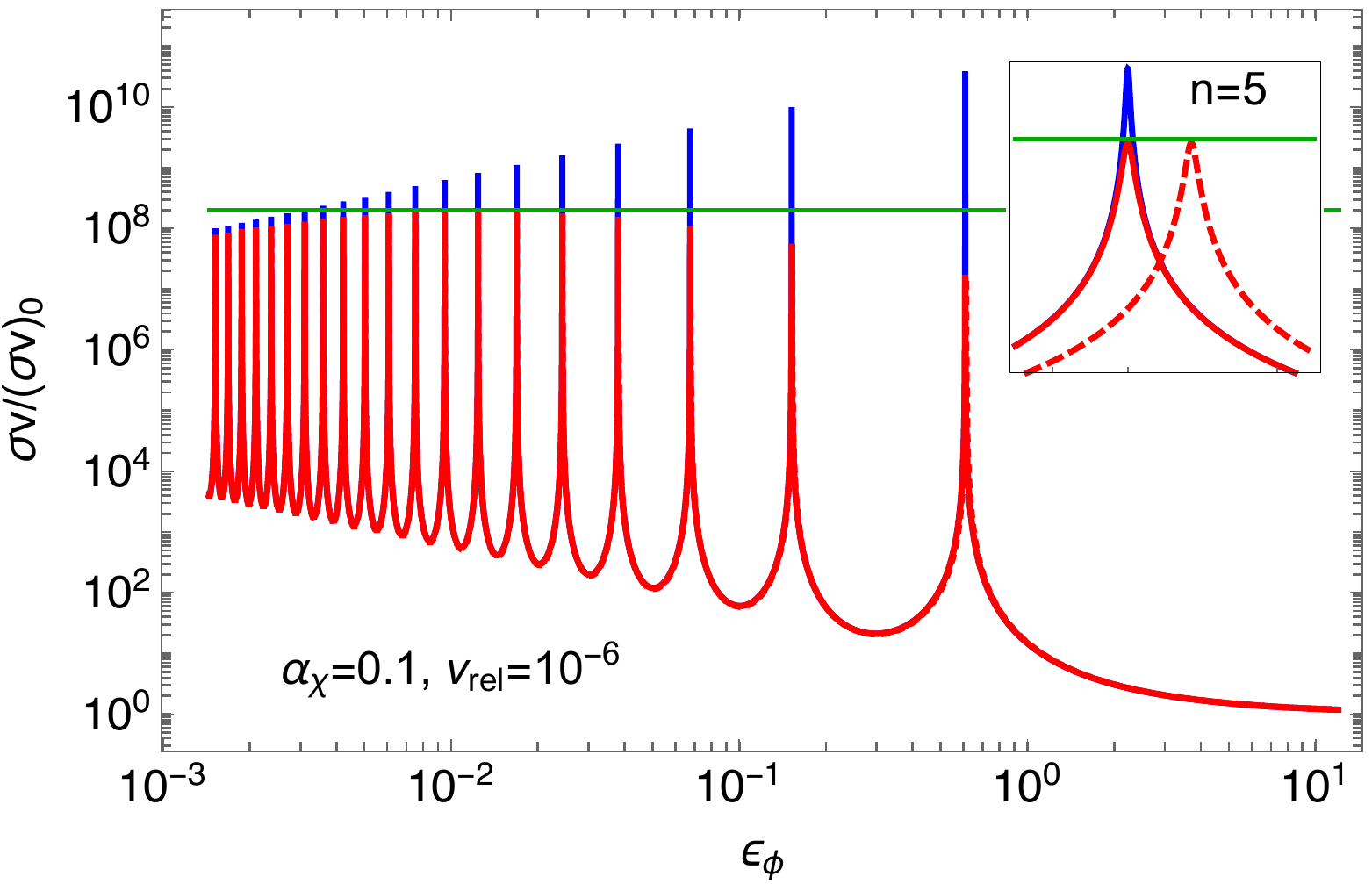}}
\caption{\emph{Left:}
Comparison between various approximations of the $s$-wave Sommerfeld factor $(\sigma v_{\text{rel}})_{\text{ann}} / (\sigma v_{\text{rel}})_{\text{ann}, 0}$ for a model with $\alpha_{\chi} \!=\! \alpha_{l} \!=\!  0.1$.
The regulated Hulth\'en potential solution (red line) with $\sigma_{\text{sc}, 0} = (\mu / 4 \pi) (\sigma v_{\text{rel}})_{\text{ann}, 0}^{2}$ is compared to the unregulated solution (blue line) on the parametric resonance point $n=5$. The black curve shows our instantaneous transitions estimate to the red solid line, which respect the $s$-wave unitarity bound shown by the green line.
For the remaining cases the exact resonance point is slightly shifted from $\epsilon_\phi = 6/(n\pi)^{2}$ and $n$ needs to be tuned to find maximal Sommerfeld enhancement:
the regulated Hulth\'en potential solution with $\sigma_{\text{sc,0}}= 3 \alpha_{\chi}^{2} \pi/(4 m_{\chi}^{2})$ for $n=5$ (dashed red) and $n=5.004$ (dotted red);
the numerical solution of the Schr\"odinger equation with a Yukawa potential \cite{Iengo:2009ni} for $n=5$ (blue dots) and $n=4.9028$ (orange dots).
\emph{Right:} Sommerfeld enhancement as a function of $\epsilon_\phi$. Same color coding as in the left panel, but only the first three entries and the unitarity bound are plotted.
The inset panel is a zoom-in around the fifth resonance, covering a 1\,\% range in $\epsilon_\phi$ and 4 orders of magnitude range in $(\sigma v_{\text{rel}})_{\text{ann}}$, showing the choice of $\sigma_{\text{sc,0}}$ only shifts the resonance slightly while leaving the maximal enhancement unaffected.}
\label{fig:Sv}
\end{figure}
\section{Derivation of analytic estimates}
\label{app:reaestimates}
In Section~\ref{app:somres} we find approximate expressions of the Sommerfeld enhancement that we then use
to estimate $\langle \sigma v_{\text{rel}} \rangle_{x^{\prime}}$.
The latter quantity is used in Section~\ref{app:softrea} to find analytical expressions for $\Gamma$, which finally allows us to estimate the change in DM abundance due to reannihilation.
Based on these results we estimate the onset time of reannihilation in Section~\ref{sec:zrea}.
\subsection{Sommerfeld enhanced annihilation on a resonance}
\label{app:somres}
In the following, we first estimate $S(v)$ and then $\langle \sigma v_{\text{rel}} \rangle_{x^{\prime}}$ in the case where the parameter $\epsilon_{\phi}$ fulfills the resonance condition.
It can be recognized from the left panel of Fig.~\ref{fig:Sv} that the Sommerfeld factor has regions where $S(v) \propto 1$, $1/v$, and  $1/v^{2}$ and a region where it starts to saturate and finally reaches a maximal value at low velocities.
We approximate the transitions between these different regions as instantaneous transitions at the following transition velocities:
\bea
v_{1/v} &=& 2 \pi \alpha_{\chi} \, ,\\
v_{1/v^{2}} &=& \frac{m_{\phi}}{m_{\chi}} \, ,\\
v_{\text{sat}} &=& \frac{\pi}{2}\frac{m_{\phi}}{m_{\chi}} \alpha_{\chi}^{2} (\alpha_{\chi} + \alpha_{l}) \, ,\\
v_{\text{max}} &=& \frac{\pi}{32} \frac{m_{\phi}}{m_{\chi}} \alpha_{\chi}^{2}(\alpha_{\chi} + \alpha_{l})\, .
\eea
We summarize this instantaneous description of $S$ in Table~\ref{tab:tab2} and demonstrate in Fig.~\ref{fig:Sv} that this approximation (black line) matches well the numerical (red solid line) solution within each definite scaling regime.
We have tested several on-resonant values of $\epsilon_{\phi}$ and found in all the cases a similarly good result.
The value of $v_{\text{sat}}$ can be obtained by equating the values of $S$ from the regime of $1/v^{2}$ scaling in Table~\ref{tab:tab2} with Eq.~\eqref{eq:smax}.
For lower velocities than $v_{\text{sat}}$ we consider that $S$ follows the scaling of the partial-wave unitarity bound, i.e., $1/v$ scaling.
$v_{\text{max}}$ is the velocity where $S$ reaches the maximal value:
\be
S^{\text{max}} = \frac{m_{\chi}}{2 \pi \alpha_{\chi} m_{\phi}} \left(\frac{4 \pi}{\mu^{2} (\sigma v_{\text{rel}})_{0}} \right)^{2},
\ee
which can directly be obtained from Eq.~\eqref{eq:svfull} with $\sigma_{\text{sc}, 0} = \frac{\mu}{4 \pi} (\sigma v_{\text{rel}})_{\text{ann}, 0}^{2}$, ignoring contributions from $T$ and taking the limit of $v\rightarrow 0$.

\begin{table}[h!]
\begin{center}
    \renewcommand{\arraystretch}{2.5}
    \begin{tabularx}{0.9\columnwidth}{ @{\hskip 3mm} l @{\hskip 3mm} | @{\hskip 5mm} c @{\hskip 5mm} || @{\hskip 3mm} l @{\hskip 3mm} | @{\hskip 5mm} c @{\hskip 5mm} }
    $v \in$ & $S(v) $  & $x^{\prime} \in$ & $\langle \sigma v_{\text{rel}} \rangle / (\sigma v_{\text{rel}})_{0} $ \\ 
    \Xhline{4\arrayrulewidth}
    $ [ \infty 	\,,\,  	v_{1/v} 			]$ 		&$\displaystyle 1$ &  $ [ 0 			\,,\,  x^{\prime}_{1/v}				]$ 		&$\displaystyle 1$\\[1.0ex] \hline
    $ [  v_{1/v}	\,,\,   v_{1/v^{2}} ]$ 			&$\displaystyle \left(\frac{v_{1/v}}{v} \right) $ &  $ [ x^{\prime}_{1/v}	 	\,,\,  x^{\prime}_{1/v^{2}}	 ]$ 			&$\displaystyle \left( \frac{x^{\prime}}{x^{\prime}_{1/v}} \right)^{1/2} $\\[1.0ex] \hline 
    $ [ 		v_{1/v^{2}}				\,,\,   v_{\text{sat}} ] $ 	&$\displaystyle \left(\frac{v_{1/v}}{v_{1/v^{2}}}\right)\left(\frac{v_{1/v^{2}}}{v}\right)^{2}$ & $ [x^{\prime}_{1/v^{2}}						\,,\,  x^{\prime}_{\text{sat}}	 ] $ 	&$\displaystyle \left( \frac{x^{\prime}_{1/v^{2}}}{x^{\prime}_{1/v}} \right)^{1/2} \left( \frac{x^{\prime}}{x^{\prime}_{1/v^{2}}} \right)$\\[1.0ex] \hline
    $ [			v_{\text{sat}}						\,,\,  v_{\text{max}} ]$ 		&$\displaystyle \left(\frac{v_{1/v}}{v_{1/v^{2}}}\right)\left(\frac{v_{1/v^{2}}}{v_{\text{sat}}}\right)^{2}\left(\frac{v_{\text{sat}}}{v}\right)$ &  $ [x^{\prime}_{\text{sat}}											\,,\, x^{\prime}_{\text{max}}]$ 		&$\displaystyle \left( \frac{x^{\prime}_{1/v^{2}}}{x^{\prime}_{1/v}} \right)^{1/2} \left( \frac{x^{\prime}_{\text{sat}}}{x^{\prime}_{1/v^{2}}} \right)\left(\frac{x^{\prime}}{x^{\prime}_{\text{sat}}} \right)^{1/2}$\\[1.0ex] \hline
     $ [			v_{\text{max}}						\,,\, 0  ]$ 		&$\displaystyle S^{\text{max}}$ &  $ [x^{\prime}_{\text{max}}											\,,\, \infty]$ 		&$\displaystyle S^{\text{max}}$
\end{tabularx}
\end{center}
\caption{Instantaneous approximation of the Sommerfeld factor and $\langle \sigma v_{\text{rel}} \rangle$ for an on-resonance $s$-wave annihilation.}
\label{tab:tab2}
\end{table}

A similar instantaneous transition description will next be used to estimate the temperature evolution of the thermally averaged cross section $\langle \sigma v_{\text{rel}} \rangle_{x'}$ defined in Eq.~\eqref{eq:sigmavaveraged}.
In the following we will drop the index $x'$ to shorten the notation.
$\langle \sigma v_{\text{rel}} \rangle$ has a definite power-law dependence on $x^{\prime}=m_{\chi}/T_{\chi}$ at temperatures where one particular scaling of $S(v)$ dominates. 
In analogy to the transition velocities, $v_{i}$, in Table~\ref{tab:tab2}, we define instantaneous transition temperatures as
\bea
x^{\prime}_{1/v}  &=&  \frac{c_{1/v}}{v_{1/v}^{2}}\, , \label{eq:xvm1} \\
x^{\prime}_{1/v^{2}}  &=&  \frac{c_{1/v^{2}}}{v_{1/v^{2}}^{2}}\, , \label{eq:xvm2}\\
x^{\prime}_{\text{sat}}  &=&  \frac{c_{\text{sat}}}{v_{\text{sat}}^{2}}\, ,\label{eq:xsatprime} \\
x^{\prime}_{\text{max}}  &=&  \frac{c_{\text{max}}}{v_{\text{max}}^{2}}\, . 
\eea
and adjust the coefficients $c_{i}$ such that the approximation coincides with the numerically obtained values of $\langle \sigma v_{\text{rel}} \rangle$ within each definite scaling regime.
For $x^{\prime}$ larger than $x^{\prime}_{\text{max}}$ we require that $\langle \sigma v_{\text{rel}} \rangle / (\sigma v_{\text{rel}})_{0} = S^{\text{max}}$, which automatically determines the last matching coefficient:
\be
c_{\text{max}} = \frac{c_{1/v} c_{1/v^{2}}}{c_{\text{sat}}} \,.
\ee
In Table~\ref{tab:tab2} we summarize the instantaneous approximation of $\langle \sigma v_{\text{rel}} \rangle$.
In particular, we find that the instantaneous approximation with $c_{1/v}=3 $, $c_{1/v^{2}}=3/2$, and $c_{\text{sat}}=1$ matches well the numerical result of $\langle \sigma v_{\text{rel}} \rangle$ within each definite scaling regime.
Next, we use this result to estimate the size of $\Gamma$.


\subsection{Estimating the maximal change in the number density}
\label{app:softrea}
In this appendix, we estimate the change in the relic abundance due to reannihilation.
The ratio between the comoving DM abundances at kinetic decoupling ($x_\text{kd}$) and today ($x_0$) can be obtained from the standard solution of the Boltzmann equation [Eq.~\eqref{eq:Yevo}], given by
\be
\frac{Y(x_{\text{kd}})}{Y(x_{0})} = 1 + \int_{x_{\text{kd}} }^{x_{0}}\text{d}x \, \frac{\Gamma}{x} \, . \label{eq:maxchange}
\ee
Here, $\Gamma$ is defined as in Eq.~\eqref{eq:gammadef}, but with the replacement $Y(x) \rightarrow Y(x_{\text{kd}})$. Note that the right hand side is thus independent of the evolution of $Y(x)$. The aim is now to further simplify this formal solution by approximating the time integral. The dominant contribution is from the $x$ range where $\Gamma$ is maximal.
In the following, we first derive simple power-law expressions of the maximal value of $\Gamma $ and second show how to approximate the time integration in various cases.

The usual order of chemical before kinetic decoupling ($x_{\text{cd}}\lesssim x_{\text{kd}}$) and that DM kinetically decouples before matter radiation equality ($x_{\text{kd}}\lesssim x_{\text{0}}$) to have an adequate structure formation history implies the following time order: $x_{\text{cd}}\lesssim x_{\text{kd}}\lesssim x_{\text{eq}}\lesssim x_{\text{0}}$, where the subscripts labels the SM photon temperature at chemical decoupling, kinetic decoupling, matter-radiation equality and today, respectively.
There are now several options to align the times $x_{1/v}$, $x_{1/v^{2}}$, and $x_{\text{sat}}$ [set by Eqs.~\eqref{eq:xvm1}, \eqref{eq:xvm2}, and \eqref{eq:xsatprime} after converting $x'$ into $x$ via Eq.~\eqref{eq:xpTOx}] in between the fixed time order $x_{\text{cd}}\lesssim x_{\text{kd}}\lesssim x_{\text{eq}}\lesssim x_{\text{0}}$.
It turns out that only five different cases (time alignments) are relevant for us and those are summarized as follows:
\begin{center}
{

\renewcommand{\arraystretch}{1.0}
    \begin{tabularx}{0.8\textwidth}	{*{8}{>{\centering\arraybackslash}X} >{\centering\arraybackslash}X}    
\arrayrulecolor{black}
Case   	& \multicolumn{2}{ |c } { $\lesssim x_{\text{cd}} \lesssim $ }	
 		& \multicolumn{2}{ c } { $\lesssim x_{\text{kd}} \lesssim $ }	
		& \multicolumn{2}{ c } { $\lesssim x_{\text{eq}} \lesssim $ } 	
		& \multicolumn{2}{ c } { $\lesssim x_{0} \lesssim $ } 
\\  
\Xhline{4\arrayrulewidth}
{\multirow{2} {*}{1.} } 
   		& \multicolumn{1}{ |c| } {}	& \multicolumn{2}{ c| } {$x_{1/v}$ }	& \multicolumn{2}{ c| }{$x_{1/v^{2}}$ }							&\multicolumn{2}{ c| }{}	& $x_{\text{sat}}$  \\ 
   		& \multicolumn{1}{ |c| } {}	& \multicolumn{2}{ c| }{}			& \multicolumn{2}{ c| } {$ x_{1/v} \lesssim  x_{1/v^{2}}$} &\multicolumn{2}{ c| }{}	& $x_{\text{sat}}$  \\ 
		\hline
{\multirow{2} {*}{2.} } 
   		& \multicolumn{1}{ |c| } {}	& \multicolumn{2}{ c| } {$x_{1/v}$ }	& \multicolumn{2}{ c| }{$x_{1/v^{2}}$ }							&\multicolumn{2}{ c| }{$x_{\text{sat}}$}	&   \\ 
   		& \multicolumn{1}{ |c| } {}	& \multicolumn{2}{ c| }{}			& \multicolumn{2}{ c| } {$x_{1/v} \lesssim  x_{1/v^{2}}$}			&\multicolumn{2}{ c| }{$x_{\text{sat}}$}	&   \\ 
\hline
{\multirow{2} {*}{3.} } 
   		& \multicolumn{1}{ |c| } {}	& \multicolumn{2}{ c| } {$x_{1/v}$ }	& \multicolumn{2}{ c| }{$x_{1/v^{2}} \lesssim x_{\text{sat}}$ }							&\multicolumn{2}{ c| }{}	&   \\ 
   		& \multicolumn{1}{ |c| } {}	& \multicolumn{2}{ c| }{}			& \multicolumn{2}{ c| } {$x_{1/v} \lesssim  x_{1/v^{2}} \lesssim x_{\text{sat}} $}			&\multicolumn{2}{ c| }{}	&   \\ 
\hline
{\multirow{2} {*}{4.} } 
   		& \multicolumn{1}{ |c| } {$x_{1/v}$}	& \multicolumn{2}{ c| } { }				& \multicolumn{2}{ c| }{$x_{1/v^{2}}$ }							&\multicolumn{2}{ c| }{$x_{\text{sat}}$}	&   \\ 
  		& \multicolumn{1}{ |c| } {$x_{1/v}$}	& \multicolumn{2}{ c| } {$x_{1/v^{2}}$}		& \multicolumn{2}{ c| }{}									&\multicolumn{2}{ c| }{$x_{\text{sat}}$}	&   \\ 
 \hline
{\multirow{2} {*}{5.} } 
   		& \multicolumn{1}{ |c| } {$x_{1/v}$}	& \multicolumn{2}{ c| } { }				& \multicolumn{2}{ c| }{$x_{1/v^{2}} \lesssim  x_{\text{sat}}$}	&\multicolumn{2}{ c| }{}	&   \\ 
  		& \multicolumn{1}{ |c| } {$x_{1/v}$}	& \multicolumn{2}{ c| } {$x_{1/v^{2}}$}		& \multicolumn{2}{ c| }{$x_{\text{sat}}$}							&\multicolumn{2}{ c| }{}	&   \\ 
\arrayrulecolor{white}
	& 	&  	&  	&  	& 	& 	 &
\end{tabularx}
}
\end{center}
\label{tab:cases}
\vspace{-2mm}
\arrayrulecolor{black}
%
The two options given in each case lead to the same result in the final form of $\Gamma$ as can be shown explicitly (without proof here).
In the first case, the Sommerfeld enhancement saturates at later times than the age of the Universe: $x_{0} \lesssim x_{\text{sat}}$.
This implies that $\Gamma$ reaches its maximal value today.
In all other cases shown in the table, the maximal value is given at the time of saturation of the Sommerfeld enhancement.
In the second case, saturation happens between matter-radiation equality and today while in the third case saturation is before matter-radiation equality.
In the fourth and fifth cases, the Sommerfeld enhancement becomes sizable at the first freeze-out as we have $x_{1/v} \lesssim x_{\text{cd}}$.

From here on we are always assuming that we are exactly on a Sommerfeld resonance point.
The maximal value of $\Gamma$ as a function of the free parameters in these five different cases can be obtain as follows.
We define $x_{\text{cd}}$ as the time when $\Gamma =1$.
Requiring $Y(x_{\text{kd}})$ to coincide  with the value of $Y$ which correspond to get the correct relic density, we can determine $x_{\text{cd}}$  as a function of $m_{\chi}$ only.
For the $m_{\chi}$ range between 10 GeV and 40 TeV we find that $x_{\text{cd}}$ varies approximately between 7 and 22.
This variation is a consequence of the fixed temperature ratio $r$ at BBN and the impact of the Sommerfeld effect on the first freeze-out temperature for DM masses above the TeV scale. 
At times later than $x_{\text{cd}}$, the evolution of $\Gamma$ in all five cases directly follows from the entries of Table~\ref{tab:tab1} and the results of the previous section.
For example, the estimate of $\Gamma_{1}$ is found to be:
\bea
\Gamma_{1} &=& \frac{(g_{s}/\sqrt{g_{\text{eff}}})_{0}}{(g_{s}/\sqrt{g_{\text{eff}}})_{\text{cd}}}  \underbrace{\frac{x_{\text{cd}}}{x_{1/v}}}_{\begin{array}{l} x_{\text{cd}}\lesssim x_{1/v} \, , \\S(v)=1\end{array}} 
\underbrace{\left(\frac{r_{1/v} x_{1/v}}{r_{\text{kd}} x_{\text{kd}}}\right)^{1/2}}_{\begin{array}{l} x_{1/v} \lesssim x_{\text{kd}},\\S(v)\propto 1/v \end{array}} 
 \underbrace{\left( \frac{r_{1/v^{2}}^{-1} }{r_{\text{kd}}^{-1}}\right)}_{\begin{array}{l} x_{\text{kd}} \lesssim x_{1/v^{2}} \, ,\\S(v)\propto 1/v \end{array}}
 \underbrace{ \left( \frac{r_{\text{eq}}^{-2}x_{\text{eq}}}{r_{1/v^{2}}^{-2}x_{1/v^{2}}} \right) }_{\begin{array}{l} x_{1/v^{2}}\lesssim x_{\text{eq}} \, , \\S(v)\propto 1/v^{2} \end{array}}
 \underbrace{ \left( \frac{r_{0}^{-2}  x_{0}^{1/2}}{r^{-2}_{\text{eq}}x^{1/2}_{\text{eq}} } \right)}_{\begin{array}{l} x_{\text{eq}} \lesssim x_{0} \, ,\\S(v)\propto 1/v^{2} \end{array}} \, .
\eea
By inserting $x_{1/v}$ and $x_{1/v^{2}}$ into this expression and applying the same procedure to the second and third cases, we find the maximum value of $\Gamma$ is given by
\bea
\Gamma_{1,2,3} = \frac{(g_{s}/\sqrt{g_{\text{eff}}})_{\text{sat}}}{(g_{s}/\sqrt{g_{\text{eff}}})_{\text{cd}}} \frac{x_{\text{cd}}}{r_{\text{sat}}^{2}}  \frac{2 \pi}{\sqrt{c_{1/v}c_{1/v^{2}}}} \frac{\alpha_{\chi} m_{\phi}}{m_{\chi}} \times \begin{cases}
\left( \frac{T^{\text{kd} \, 2}_{l}}{T^{\text{eq}}_{\gamma} T^{0}_{\gamma}} \right)^{1/2} & \text{ for }T^{0}_{\gamma} \gtrsim T^{\text{sat}}_{\gamma}\, ,\\  
\left( \frac{T^{\text{kd} \, 2}_{l}}{T^{\text{eq}}_{\gamma} T^{\text{sat}}_{\gamma}} \right)^{1/2} & \text{ for } T^{\text{eq}}_{\gamma} \gtrsim T^{\text{sat}}_{\gamma} \gtrsim T^{0}_{\gamma}\, , \\  
\left( \frac{T^{\text{kd}}_{l}}{ T^{\text{sat}}_{\gamma}} \right) & \text{ for } T^{\text{sat}}_{\gamma} \gtrsim T^{\text{eq}}_{\gamma} \, .
\end{cases} 
\eea
and in the last two cases where $x_{\text{cd}} \gtrsim x_{1/v}$ we find
\bea
\Gamma_{4,5}  = \frac{(g_{s}/\sqrt{g_{\text{eff}}})_{\text{sat}}}{(g_{s}/\sqrt{g_{\text{eff}}})_{\text{cd}}}  \frac{(x_{\text{cd}} r_{\text{cd}})^{1/2}}{r_{\text{sat}}^{2}} \frac{1}{\sqrt{c_{1/v^{2}}}} \frac{ m_{\phi}}{m_{\chi}} \times \begin{cases} 
\left( \frac{T^{\text{kd} \, 2}_{l}}{T^{\text{eq}}_{\gamma} T^{\text{sat}}_{\gamma}} \right)^{1/2} & \text{ for } T^{\text{eq}}_{\gamma} \gtrsim T^{\text{sat}}_{\gamma} \gtrsim T^{0}_{\gamma}\, , \\  
\left( \frac{T^{\text{kd}}_{l}}{ T^{\text{sat}}_{\gamma}} \right) & \text{ for } T^{\text{sat}}_{\gamma} \gtrsim T^{\text{eq}}_{\gamma} \, .
\end{cases} 
\eea
The kinetic decoupling temperature in the equal charge case $(g_{\chi} =g_{l})$ and two species of $l$ (particle and anti particles) is given by \cite{Binder:2016pnr}
\be
\frac{T^{\text{kd}}_{l}}{1 \, \text{keV}} = 0.25 \times \left(\frac{r_{\text{kd}}}{0.36}\right)^{-1/2} \left( \frac{\alpha_{\chi}}{0.025}\right)^{-1/2} \left( \frac{m_{\chi}}{ \text{1 TeV}} \right)^{1/4}\left( \frac{m_{\phi}}{ \text{1 MeV}} \right) \,.
\label{eq:kd}
\ee
We find the saturation temperature from Eq.~\eqref{eq:xsatprime}:
\bea
T^{\text{sat}}_{\gamma} &=& \frac{\pi}{ r_{\text{sat}} \sqrt{c_{\text{sat}}}} \alpha_{\chi}^{3} m_{\phi} (x_{\text{kd}}^{l})^{-1/2} \\
&=& 2.96 \times 10^{-3} \, \text{eV} \left(\frac{r_{\text{sat}}}{0.36}\right)^{-1}\left( \frac{\alpha_{\chi}}{0.025} \right)^{3} \left( \frac{m_{\chi}}{ \text{1 TeV}} \right)^{-1/2} \left( \frac{m_{\phi}}{ \text{1 MeV}} \right)\left( \frac{T^{\text{kd}}_{l}}{ \text{1 keV}} \right)^{1/2} \label{eq:saturationtemperature} \, .
\eea

We now simplify the time integration of $\Gamma$ to obtain the change in DM abundance due to reannhiation.
A simple case is the case 1 where $x_{0} \lesssim x_{\text{sat}}$.
$\Gamma$ takes the maximal value of $\Gamma_{1}$ today and hence the integration can be simplified as
\be
\int_{x_{\text{kd}} }^{x_{0}}\text{d}x \,  \frac{\Gamma}{x}  \approx \int_{x_{\text{eq}} }^{x_{0}}\text{d}x \,  \frac{\Gamma}{x} \simeq 2 \times \Gamma_{1} \, .
\ee
Inserting this result into Eq.~\eqref{eq:maxchange} and solving for $m_{\chi}$ for given $m_{\phi}$ we find the maximal DM changes that reannihilation can cause. 
By setting the left hand side of Eq.~\eqref{eq:maxchange} to 1.01 and 1.1 (correspond to ``max 1\%'' and ``max 10\%'' DM changes, respectively) we obtain the most left parts of the red lines in Fig.~\ref{fig:deg}.
In practice, this equation is solved numerically since we use tabulated values for $x_{\text{cd}}$ and $\alpha_{\chi}$, where the latter quantity is chosen such that $Y(x_{\text{kd}})$ gives the correct relic density.
Another simple case is the the case 5 where $x_{\text{sat}} \lesssim x_{\text{eq}}$.
Here, the maximum value of $\Gamma$ is given by the saturation temperature in the radiation dominated epoch, leading to the simplification:
\be
\int_{x_{\text{kd}} }^{x_{0}}\text{d}x \,  \frac{\Gamma}{x}  \approx \int_{x_{1/v^{2}}}^{x_{\text{sat}}} + \int_{x_{\text{sat}}}^{x_{\text{max}}} + \int_{x_{\text{max}}}^{x_{\text{eq}}} \text{d}x \,  \frac{\Gamma}{x} \approx \Gamma_{4} \left[2 + \log\left(T^{\text{sat}}_{\gamma} / T^{\text{max}}_{\gamma} \right) \right] \, .
\ee
In the last approximation we assumed that $T^{\text{sat}}_{\gamma} \ll T^{1/v^{2}}_{\gamma}$ and $T^{\text{max}}_{\gamma} \gg T^{\text{eq}}_{\gamma}$.
The temperature ratio appearing in the latter equation is a constant and given by $T^{\text{sat}}_{\gamma}/T^{\text{max}}_{\gamma} \simeq 34$, which can be seen by applying the definitions.
The abundance ratio has a power-law dependence on the parameters and corresponds to the segments of the red lines in the top right part of Fig.~\ref{fig:deg}.
For the intermediate regimes where saturation happens close to today or to matter-radiation equality, a simple power-law scaling cannot be found for capturing accurately the transitions.
These regimes are the regions in Fig.~\ref{fig:deg} where the red curves start to bend in the $\log(m_{\chi})$-$\log(m_{\phi})$ plane.
The procedure to obtain the solution in these regimes are still the same as in the most simple cases described above, however, the expressions become lengthy and for simplicity we do not show these cases here.
Note that in all the estimates of $\Gamma$ shown here, we have neglected the minor impact of the dark energy as well as the effect of non-linear structure formation, assuming our homogeneous DM density treatment is valid until today.
To evaluate our estimates the following values are used:
\bea
r_{0} &=& 0.36 \, , \\
T^{\text{eq}}_{\gamma} &=& 0.80 \, \text{eV} \, , \\
T^{0}_{\gamma} &=& 2.34 \times 10^{-4} \, \text{eV} \, , \\
(g_{s} / \sqrt{g_{\text{eff}}})_{0} &=& 2.12 \, .
\eea

\subsection{Redshift of reannihilation onset}
\label{sec:zrea}
In Section~\ref{sec:estimates}, we have defined the onset of reannihilation as the redshift $z_{\text{rea}}$ where the comoving number density changes first by $1 \, \%$ after kinetic decoupling.
Using Eq.~\eqref{eq:maxchange}, $z_{\text{rea}}$ can be found by solving the integral equation
\be
0.01= \int_{x_{\text{kd}}}^{x_{\text{rea}}} \text{d}x \,  \frac{\Gamma}{x} \,, \label{eq:intzrea}
\ee
where $x_{\text{rea}}=\frac{m_{\chi}}{T_{\gamma}^{0}(1+z_{\text{rea}})}$.
We are mainly interested in the case where reannihilation happens between recombination and today.
It turns out that the relevant parameter region is where $x_{\text{cd}} \lesssim x_{1/v}$ and $\Gamma$ as a function of temperature can be obtained from $\Gamma_{1/2/3}$: 
\bea
\Gamma(x) = 0.033 \times \frac{9.7}{(g_{s} / \sqrt{g_{\text{eff}}})_{\text{cd}}} \frac{x_{\text{cd}}}{18} \left( \frac{\alpha_{\chi}}{0.02} \right) \left( \frac{m_{\chi}}{\text{TeV}} \right)^{-1}\left( \frac{m_{\phi}}{ \text{MeV}} \right) \left( \frac{T^{\text{kd}}_{l}}{0.25 \, \text{keV}} \right) \times
 \begin{cases} 
\left( \frac{x}{x_{0}} \right)^{1/2}  & x \gtrsim x_{\text{eq}} \, , \\  
\left( \frac{x_{\text{eq}} }{x_{0}}\right)^{1/2}  \frac{x}{x_{\text{eq}}}  &  x_{\text{eq}}  \gtrsim x \, .
\end{cases}
\eea
The time integration can be approximated as:
\be
\int_{x_{\text{kd}}}^{x_{\text{rea}}} \text{d}x \, \frac{\Gamma}{x} \approx \int_{x_{1/v^{2}}}^{x_{\text{eq}}} + \int_{x_{\text{eq}}}^{x_{\text{rea}}} \text{d}x \,  \frac{\Gamma}{x} \approx \Gamma(x_{\text{rea}}) \left[ 2 - \left(T^{0}_{\gamma} / T^{\text{eq}}_{\gamma}\right)^{1/2} (1 + z_{\text{rea}})^{1/2} \right] \,.
\ee
Taking this approximation in Eq.~\eqref{eq:intzrea} and solving for fixed $z_{\text{rea}}$ we obtain the green lines in Fig.~\ref{fig:deg}.
In the case where $x_{\text{rea}} \gg x_{\text{eq}}$ we can approximate the integral as:
\be
\int_{x_{\text{kd}}}^{x_{\text{rea}}} \text{d}x \,  \frac{\Gamma}{x} \approx 2 \Gamma(x_{\text{rea}}) \,.
\ee
Taking this approximation in Eq.~\eqref{eq:intzrea} and solving for $z_{\text{rea}}$ we finally obtain Eq.~\eqref{eq:zrea}.

In regions where reannihilation can only change  the DM abundances by less than $1 \, \%$, $z_{\text{rea}}$ is no longer defined ---  in Fig.~\ref{fig:deg} this is where the green lines stop.
Note that we have implicitly assumed that the saturation temperature is much lower than the reannihilation temperature. In the critical region, where the saturation redshift approaches $z_{\text{rea}}$ this approximation is no longer valid, and we indicate this by the solid green curves changing into  dashed green curves in Fig.~\ref{fig:deg}. Since the dashed region is outside the SIDM blue band we do not investigate this case further, but we have confirmed that our numerical code exactly reproduce our estimates in its valid regime but starts to deviate when the green lines becomes dashed.



\section{Standard Hubble expansion rate}
\label{sec:stdhubble}
The Hubble expansion rate as a function of the standard energy densities is given by Eq.~\eqref{eq:hlcdm}.
When including reannihilation we replace $\rho_{c}$ by $\rho_{\text{dark}}$ via Eq.~\eqref{eq:rhoDM} and when studying decaying DM we replace $\rho_{c}$ via Eq.~\eqref{eq:endectotal}.
In all cases, we take an effective neutrino mass $m_\nu$ into account in the time evolution of $\rho_{\nu}$.
We introduce a single massive eigenstate (minimum-mass normal hierarchy) such that the SM neutrino energy density evolves according to
\be
\frac{\rho_{\nu}}{\rho_{\gamma}} = \frac{ N_{\text{eff}}}{3} \frac{7}{8} \left( \frac{4}{11} \right)^{4/3} \left[2 +  \frac{I_{\nu} \left(\frac{m_{\nu} }{T_{\nu}^{0} (1+z)}\right)}{I_{\nu} (0)} \right] \,,
\ee
where $T_{\nu}^{0} = (N_{\text{eff}}/3)^{1/4}(4 / 11)^{1/3} T^{0}_{\gamma}$ and
\be
I_{\nu} (x) = \frac{1}{\pi^{2}} \int^{\infty}_{0} dy \sqrt{x^{2} + y^{2}} \frac{y^{2}}{e^{y} + 1} \,,
\ee
with $I_{\nu} (0) = 7 \pi^{2} / 120$ and the default value of the CMB temperature of today is $T^{0}_{\gamma} = 2.7255 \pm 0.0006$\,K~\cite{Fixsen:2009ug}.
We derive the photon energy density from the temperature of today to be
\be
\Omega_{\gamma}h^{2} = 2.4728 \times 10^{-5} \label{eq:photonrelic} \,
\ee
and other default parameters that we use, from Planck 2015~\cite{Ade:2015xua}, are 
\bea
m_{\nu} &=& 0.06 \, \text{eV} \, ,\\
N_{\text{eff}} &=& 3.046 \, .
\eea
Furthermore, we use the results of the Planck 2015 (TT+lowP) analysis~\cite{Ade:2015xua} where the relevant base parameters are constrained to be
\bea
\Omega_{c} h^{2} &=& 0.1197 \pm 0.0022 \, , \\
\Omega_{b} h^{2} &=& 0.02222 \pm 0.00023 \, ,
\eea
and the derived parameters from the same analysis are given by
\bea
\Omega_{\Lambda} &=& 0.685 \pm 0.013 \, , \\
h &=& 0.6731 \pm 0.0096 \, , \label{eq:h0}\\
z_{*} &=& 1090.09 \pm 0.42 \, ,\\
z_{\text{drag}} &=& 1059.57 \pm 0.46 \label{eq:zdragplanck} \, .
\eea
Using Eqs.~\eqref{eq:photonrelic}--\eqref{eq:zdragplanck} in Eqs.~\eqref{eq:thstar}--\eqref{eq:R}, we
reproduce the Planck 2015 reported values (given within the parenthesis below) of $100 \theta_{*}$, $r_{s} (z_{*})$ and $r_{s} (z_{\text{drag}})$:
\bea
100 \theta_{*} &=& 1.04103 ~ (1.04105 \pm 0.00046) \, ,\\
r_{s} (z_{*}) &=& 144.625 ~ (144.61 \pm 0.49) \, , \\
r_{s} (z_{\text{drag}}) &=& 147.34 ~ (147.33 \pm 0.49) \,.
\eea

\twocolumngrid
\bibliography{SIDM}

\end{document}